\documentclass[acmsmall,screen]{acmart}

\usepackage[table]{xcolor}
\usepackage{makecell}
\usepackage{balance}
\usepackage[utf8]{inputenc}
\usepackage{tcolorbox}
\usepackage{enumitem}
\usepackage{arydshln}
\usepackage{multirow}
\usepackage{subcaption}
\usepackage{array}    
\usepackage{float}
\usepackage{parselines} 
\usepackage{listings}
\usepackage{wrapfig}
\usepackage{longtable}
\usepackage{multirow,makecell,graphicx}
\usepackage[table]{xcolor}  
\usepackage{colortbl}
\usepackage{pifont}         

\usepackage{color}

\usepackage{graphicx}
\usepackage{pifont}

\definecolor{customblue}{HTML}{006ca6}
\definecolor{customgreen}{HTML}{009264}
\definecolor{custombrown}{HTML}{ff3d00}
\AtEndPreamble{
 \usepackage{hyperref}
 \hypersetup{
  colorlinks = true,
  linkcolor = customblue,
  anchorcolor = customblue,
  citecolor = customgreen,
  filecolor = customblue,
  urlcolor = customblue
 }
}

\definecolor{darkgreen}{rgb}{0,0.5,0}

\lstdefinelanguage{TypeScript}{
  keywords={const, let, var, function, if, else, return, async, await, import, from, export, default, class, extends, constructor, super, this, public, private, protected, static, get, set, new, instanceof, typeof, void, number, string, boolean, any, unknown, never, interface, implements, enum, type, declare, namespace, module, as, in, of, for, while, do, switch, case, break, continue, throw, try, catch, finally, null, undefined, true, false, Promise, Array, Object, Symbol, Map, Set, Date, RegExp, Error, Math, JSON, console},
  sensitive=true,
  morecomment=[l]{//},
  morecomment=[s]{/*}{*/},
  morestring=[b]",
  morestring=[b]',
  morestring=[b]`,
  ndkeywords={true, false, null, undefined, void, number, string, boolean, any, unknown, never, Promise, Array, Object, Symbol, Map, Set, Date, RegExp, Error, Math, JSON, console},
  keywordstyle=\color{blue}\bfseries,
  ndkeywordstyle=\color{red}\bfseries,
  identifierstyle=\color{black},
  commentstyle=\color{gray}\itshape,
  stringstyle=\color{darkgreen}\itshape,
  basicstyle=\ttfamily\scriptsize, 
  numberstyle=\tiny,
  stepnumber=1,
  numbersep=-2pt,
  showstringspaces=false,
  breaklines=true,
  captionpos=b,
}

\AtBeginDocument{%
  \providecommand\BibTeX{{%
    \normalfont B\kern-0.5em{\scshape i\kern-0.25em b}\kern-0.8em\TeX}}}

\newcommand{\find}[1]{
\begin{tcolorbox}[leftrule=1mm,toprule=0mm,bottomrule=0mm,left=1pt,right=2pt,top=2pt,bottom=2pt
]
\em #1
\end{tcolorbox}
}

\lstdefinestyle{JAVA}{
  language=JAVA,
  moredelim=[is][\underbar]{_}{_},
}

\begin{document}

\title{A Comparative Study of Android Performance Issues in Real-world Applications and Literature}

\author{Dianshu Liao}
\affiliation{
  \institution{The Australian National University}
  \city{Canberra}
  \country{Australia}}
\email{dianshu.liao@anu.edu.au}

\author{Shidong Pan}
\affiliation{
  \institution{CSIRO's Data61 \& The Australian National University}
  \city{Canberra}
  \country{Australia}}
\email{Shidong.pan@anu.edu.au}

\author{Siyuan Yang}
\affiliation{
  \institution{The Australian National University}
  \city{Canberra}
  \country{Australia}}
\email{u7439286@anu.edu.au}

\author{Yanjie Zhao}
\affiliation{
  \institution{Huazhong University of Science and Technology}
  \city{Wuhan}
  \country{China}}
\email{yanjie_zhao@hust.edu.cn}

\author{Zhenchang Xing}
\affiliation{
  \institution{CSIRO's Data61 \& Australian National University}
  \city{Canberra}
  \country{Australia}}
\email{zhenchang.xing@data61.csiro.au}

\author{Xiaoyu Sun}
\affiliation{
  \institution{The Australian National University}
  \city{Canberra}
  \country{Australia}
  }
\authornote{Xiaoyu Sun is the corresponding author.}
\email{Xiaoyu.Sun1@anu.edu.au}

\begin{abstract}

Performance issues in Android applications significantly undermine users' experience, engagement, and retention, which is a long-lasting research topic in academia.
Unlike functionality issues, performance issues are more difficult to diagnose and resolve due to their complex root causes, which often emerge only under specific conditions or payloads. 
Although many efforts have attempted to mitigate the impact of performance issues by developing methods to automatically identify and resolve them, it remains unclear if this objective has been fulfilled, and the existing approaches indeed targeted on the most critical performance issues encountered in real-world settings.
To this end, we conduct a large-scale comparative study of Android performance issues in real-world applications and literature. 
Specifically, we started by investigating real-world performance issues, their underlying root causes (i.e., contributing factors), and common code patterns. 
We then took an additional step to empirically summarize existing approaches and datasets through a literature review, assessing how well academic research reflects the real-world challenges faced by developers and users. Our comparison results show a substantial divergence exists in
the primary performance concerns of researchers, developers, and users. 
Among all the identified factors, 57.14\% have not been examined in academic research, while a substantial 63.41\% remain unaddressed by existing tools, and 70.73\% lack corresponding datasets. 
This stark contrast underscores a substantial gap in our understanding and management of performance issues. Consequently, it is crucial for our community to intensify efforts to bridge these gaps and achieve comprehensive detection and resolution of performance issues.

\end{abstract}



\maketitle

\section{Introduction}\label{sec:Intro}

With the increasing popularity of smartphones, the Android system has become one of the largest mobile platforms in the world, holding a 70.69\% market share worldwide~\cite{iPhoneAndroidStatistics}. 
The exponential growth of Android usage has seen an unprecedented expansion within the mobile app industry, with millions competing for users' attention. 
In such a fiercely competitive circumstance, developers face pressure to create new, fancy apps in the quickest manner~\cite{Nikzad2014APEAA}. 
However, this emphasis on speed often leads to a focus on functional requirements, while non-functional requirements, particularly performance, are overlooked—even though they are crucial to the overall quality of the apps~\cite{Hort2021ASO}.

According to the classic SE textbook from Ian Sommerville, performance issues arise when software meets functional requirements but operates inefficiently or wastes resources, which can negatively affect its reliability~\cite{sommerville2011software}. 
These issues are particularly critical in Android mobile devices, where resources like battery life, CPU, memory, and storage are limited~\cite{Guo2013CharacterizingAD, Liu2014CharacterizingAD, Rua2023ALE, Lee2019ImprovingEE}.
In Android apps, performance issues affecting efficiency, resource utilization, or app smoothness can drastically reduce user satisfaction and software reliability~\cite{Li2019DetectingAD, Liu2014CharacterizingAD, Gao2017EveryPC, Liu2018NavyDroidAE}. 
These issues include, but are not limited to, excessive energy consumption~\cite{Liu2014CharacterizingAD, Banerjee2018EnergyPatchRR, Pathak2012WhatIK, Carette2017InvestigatingTE}, redundant memory consumption~\cite{Guo2013CharacterizingAD, Song2021IMGDroidDI, Habchi2019TheRO}, and responsiveness delays~\cite{Farooq2018RuntimeDroidRR, Kang2016DiagDroidAP, Lin2024AgingOG}.

However, due to the non-functional nature of performance issues, their problems do not immediately manifest. 
Additionally, performance issues can be caused by multiple factors, requiring in-depth analysis and debugging to diagnose and identify the root causes~\cite{Liu2016FixingRL, Habchi2018OnAL, Jabbarvand2019SearchBasedET, Li2019CharacterizingAD, Liu2014GreenDroidAD}. 
Furthermore, some performance issues only appear under specific usage scenarios or loads~\cite{Behrouz2016EnergyawareTM, Jabbarvand2019SearchBasedET, Yan2013SystematicTF, Farooq2018RuntimeDroidRR}.
Consequently, manually identifying performance issues is time-consuming and labor-intensive, making it difficult for developers to address all problems during the coding and testing phase.
Therefore, it is essential for researchers to support developers in analyzing, detecting, and resolving Android performance issues~\cite{Habchi2018OnAL, Li2019DetectingAD, Li2022CombattingEI, Vsquez2015HowDD}.

To address performance issues in Android applications, researchers have explored the topic from multiple perspectives. 
On one hand, several studies aim to understand real-world performance issues by analyzing practical data sources~\cite{Kumari2024AnES, Noor2022EndUP, Das2016AQA, Vsquez2015HowDD}.
For instance, Noor et al.~\cite{Noor2022EndUP} examined 368,704 low-rated Google Play reviews and identified eight categories of user-reported performance issues, such as responsiveness delays and graphical glitches. 
Kumari et al.~\cite{Kumari2024AnES} analyzed 385 Stack Overflow posts to summarize developer-discussed performance issues and their causes, including UI/UX inefficiencies and problematic development practices.
On the other hand, academic research has focused on performance-related code smells and design flaws~\cite{LeGoaer2022ecoCodeAS, Wu2023ASL, Fawad2024AndroidSC, Palomba2019OnTI}. 
For example, Wu et al.~\cite{Wu2023ASL} reviewed 35 studies and categorized Android code smells into five types, including energy inefficiency and memory overuse. 
Similarly, Fawad et al.~\cite{Fawad2024AndroidSC} examined 79 papers and identified 237 Android code smell types, many of which are closely related to performance concerns, such as wakelock misuse and inefficient database queries.
In addition to identifying performance issues, researchers have proposed various techniques to mitigate them~\cite{Li2019CharacterizingAD, Banerjee2014DetectingEB, Song2021IMGDroidDI, Rua2023ALE, Das2020CharacterizingTE, Vsquez2015HowDD, Hort2021ASO, Liu2019DroidLeaksAC, Habchi2019TheRO}. 
For example, in the area of static analysis, Li et al.~\cite{Li2019CharacterizingAD} developed TAPIR, which analyzes anti-patterns in app call graphs to detect inefficient image displays that can lead to memory consumption, energy consumption, and responsiveness issues. 
Meanwhile, dynamic approaches such as \textmu Droid~\cite{Jabbarvand2017DroidAE} have been proposed to detect energy waste caused by resource underutilization, wakelock misuse, and inefficient loops by monitoring abnormal energy consumption.

However, despite extensive research efforts, the study of Android performance issues remains fragmented.
Most existing studies focus on a single data source or perspective, such as the user perspective from Google Play~\cite{Noor2022EndUP}, the developer perspective from GitHub issues~\cite{Das2016AQA, Vsquez2015HowDD, Liu2014CharacterizingAD}, or the researcher perspective from academic literature~\cite{Wu2023ASL, Fawad2024AndroidSC, Hort2021ASO}.
Moreover, these studies examine performance issues from different dimensions: some focus on symptoms (e.g., memory leaks~\cite{Banerjee2018EnergyPatchRR, Liu2016FixingRL}), others on root causes (e.g., improper wakelock management~\cite{Pathak2012WhatIK}), or specific code-level patterns (e.g., inefficient image loading~\cite{Song2021IMGDroidDI}).
Terminological inconsistencies further exacerbate this fragmentation, as similar issues are defined differently across studies, such as \textit{green bugs}~\cite{Goar2020EnforcingGC}, \textit{energy leaks}~\cite{Carette2017InvestigatingTE}, and \textit{energy smells}~\cite{Iannone2020RefactoringAE}.
As a result, it remains unclear what the current state of Android app performance analysis is, whether current research is focused on key performance issues encountered in real-world settings, and to what extent existing methods effectively and comprehensively address these critical real-world challenges.
Specifically, in this work, we formulate these concerns into two research questions (RQs) that we aim to answer through empirical evidence and explanatory results. The two RQs are summarized as follows.

\textbf{RQ1:} What are the primary performance issues that developers and users are most concerned with in the real-world settings \textbf{(RQ1.1)}? And what are the root causes \textbf{(RQ1.2)} and common code patterns \textbf{(RQ1.3)} of these issues?

\textbf{RQ2:} Do the performance issues investigated in academic research align with those that developers and users are most concerned with in real-world \textbf{(RQ2.1)}? To what extent do the tools developed in academic research address real-world performance issues \textbf{(RQ2.2)}? Are the datasets developed in academic research adequate for evaluating the effectiveness of these tools in tackling Android performance issues \textbf{(RQ2.3)}?

To answer these RQs, we began by conducting a large-scale analysis of real-world discussions on Android performance issues.
Specifically, we collected 60,684 negative reviews from Google Play, 749,067 question threads from Stack Overflow, and 16,977 issues with 344,922 commits from GitHub to identify performance issues experienced by users and developers, analyze their root causes (i.e., contributing factors), and examine common code patterns.
Following this, we conducted a systematic literature review
on Android performance analysis to investigate whether existing works indeed touch on the leading performance issues encountered in real-world settings.
Through a comparison of academic findings with real-world data, we developed a comprehensive taxonomy covering seven categories of performance issues and 82 contributing factors (Table~\ref{tab:performance__issues}). 
This taxonomy (1) unifies inconsistent terminologies that fragmented prior studies, (2) explicitly distinguishes consequences from causes, providing a clearer mapping between real-world problems and their contributing factors, and (3) integrates user, developer, and researcher perspectives within a single framework. 
Together, these features enable systematic cross-perspective analysis and allow us to identify substantial gaps in current academic research on Android performance analysis.
Specifically, it turns out there is a substantial divergence in primary concerns among researchers, developers, and users.
Additionally, it is surprising to find that among all identified factors, 57.14\% have not been examined in academic research, while a substantial 63.41\% remain unaddressed by existing tools, and 70.73\% lack corresponding datasets.
On top of these findings, we further propose future research directions to bridge these gaps.

\noindent\textbf{In this paper, we made the following key contributions:}
\begin{itemize} [leftmargin=*]
    \item \textbf{Cross-perspective analysis of real-world and academic sources.}
    We collected and analyzed performance-related data from Google Play (user perspective), Stack Overflow and GitHub (developer perspective), and academic literature (researcher perspective), providing the first unified investigation across real-world and academic sources from multiple stakeholder perspectives.
    \item \textbf{A unified consequence–cause taxonomy.}  
    We proposed a hierarchical taxonomy that systematically links observable performance issues (e.g., sluggish UI, memory bloat) to their contributing factors (e.g., inefficient image handling, unreleased wakelocks). The taxonomy covers seven types of performance issues and identifies 82 contributing factors, enabling consistent classification and comparison across sources.
    \item \textbf{Quantitative gap analysis between research and practice.}  
    Using the taxonomy, we compared literature with real-world data to assess how well current research, tools, and datasets align with real-world concerns, and proposed directions to address the identified gaps.
    \item \textbf{Artifact release for reproducibility.}
    We have released our artifacts~\cite{our_repo}, including source code, dataset, and summarized results, to support future research and community development.
\end{itemize}
\section{A Large-scale Exploration of Real-world Discussions on Android Performance Issues} \label{sec_realworld_study}

While performance issues frequently arise in real-world scenarios, it remains unclear which specific issues concern users and developers, along with their root causes and common code patterns.
To address this, we conducted an exploratory study to collect, identify, and analyze real-world performance issues as discussed on popular online platforms.

\begin{figure*}
    \centering
    \setlength{\belowcaptionskip}{-10pt}
    \includegraphics[width=0.98\textwidth]{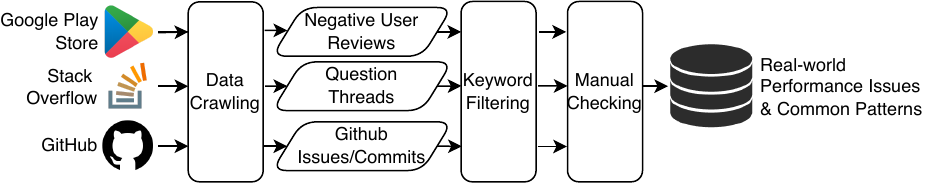}
    \caption{The process of real-world data collection.}
    \label{fig:data collection}
\end{figure*}

\subsection{Data Collection} \label{sec:data collection}

Following common practices in analyzing large-scale online discussions~\cite{Guo2013CharacterizingAD, Liu2014CharacterizingAD, Rua2023ALE, Banerjee2018EnergyPatchRR}, we collected data from three popular crowd-sourced platforms: Google Play, Stack Overflow, and GitHub.
Indeed, Google Play provides user feedback to assess the impact on end-users, Stack Overflow offers discussions that reveal common issues and solutions, and GitHub contains issues and commit histories that reflect the challenges developers encounter.
Fig.~\ref{fig:data collection} shows our three-step pipeline to perform the real-world exploration.
Briefly, we first crawl user reviews from the Google Play Store, question threads from Stack Overflow, and issue-commit pairs from GitHub. 
Then, we filtered this data using performance-related keywords.
Lastly, we manually checked real-world performance issues, root causes, and their common code patterns.

\subsubsection{Data Crawling} \label{sec:data crawling}
Here, we elaborate the details of data crawling as follows.

\noindent \textbf{[Google Play - User Reviews.]}
To collect user reviews, we focus on those from the Google Play Store, the largest Android app marketplace globally. 
This platform plays a crucial role in the Android ecosystem, enabling app distribution and discovery for millions of users worldwide. Users often report performance issues through their reviews on the Google Play Store, providing valuable insights for researchers~\cite{Liu2014CharacterizingAD}.
We collected user reviews from the apps on Google Play based on the \href{https://www.appbrain.com/stats/google-play-rankings}{\emph{AppBrain}} popularity ranking.
To ensure representativeness, we collected the top 200 apps in five monetary categories (top free, top paid, top grossing, top new free, top new paid) across 30 countries, totaling 7,681 unique apps.
We used the Google Play scraper library\footnote{google-play-scraper:~\url{https://pypi.org/project/google-play-scraper/}} and successfully collected 909,430 user reviews from these apps.
Please note that we excluded user reviews that only have a rating score (1-5) without comments.
Intuitively, performance issues are embedded in the negative user reviews, so we used the sentiment analysis model SiEBERT~\cite{hartmann2023} to identify negative reviews. This model is a fine-tuned checkpoint of RoBERTa~\cite{Liu2019RoBERTaAR}, with fine-tuning datasets from diverse text sources such as reviews and tweets.
As a result, we obtained 60,684 negative user reviews for further processing.

\noindent \textbf{[Stack Overflow - Question Threads.]}
We also collected performance-related question threads from Stack Overflow, a widely recognized Q\&A platform for programmers and developers, which plays a crucial role in the Android ecosystem by facilitating knowledge sharing, problem-solving, and community support~\cite{Wu2022RetrievingAK,Huang2018APIMR}. 
Specifically, we explored the performance issues developers posted and discussed in the development process.
Each question thread on Stack Overflow typically reflects a real-world programming challenge, with the follow-up discussions offering additional insights, perspectives, and potential solutions.
To understand developers' perspectives on Android performance issues, we crawled all question threads labeled with ``android'' with code, resulting in 749,067 questions.

\noindent \textbf{[Github - Issues \& Commits.]}
Additionally, we collected performance issues and commits for open-source apps from GitHub, the most widely used platform for open-source development. GitHub issues provide structured information of reported performance issues, while commits create a clear and traceable history of code fixes made as solutions. Consequently, both issues and commits offer valuable insights into how developers effectively resolve these issues during the maintenance process. To access each app's corresponding GitHub repository, we utilized F-Droid~\cite{FDroid}, an open-source Android app repository. We chose F-Droid because it facilitates easier inspection of code patterns in open-source applications compared to closed-source ones. 
To gather a representative sample of Android apps, we crawled 17 categories of open-source Android apps along with their GitHub links from F-Droid, resulting in an initial dataset of 4,228 open-source Android apps.
To ensure that the selected apps are not dummy projects, we introduced an additional validation step. Specifically, we verified whether each app also had a corresponding link to the Google Play Store. This yielded 1,643 apps that were both open-source and publicly distributed.
From the repositories of these 1,643 apps, we extracted a total of 16,977 GitHub issues and 344,922 commits for our subsequent analysis.

\subsubsection{Keyword Filtering} \label{sec_keyword_filtering}
As we focus on Android performance issues, we implemented a keyword-based filtering approach to select relevant data.
To compile an accurate keyword list, we conducted a Google search for ``Android performance'', reviewing the top 10 most relevant pages, which included 100 websites.
Additionally, we examined several foundational papers on Android performance issues~\cite{Hort2021ASO, Liu2014CharacterizingAD, Habchi2019TheRO, Rua2023ALE, Das2020CharacterizingTE, Das2016AQA}.
From these sources, we obtained a list of 87 keywords about Android performance issues.
The keyword list includes terms like \textit{``slow response''}, \textit{``memory leak''}, \textit{``battery drain''}, \textit{``cpu utilization''}, \textit{``app size''} and so on.
The full list of keywords is available in~\cite{our_repo}.
These keywords were used to match the titles of user reviews, Stack Overflow posts, GitHub issues, and commits.
It is important to note that title matching served only as an initial filtering step.
In subsequent manual analysis, we further examined the associated descriptions and code snippets to extract actual performance issues and discover contributing factors, which may not have been explicitly mentioned in the original keyword list.
This deeper analysis allowed us to go beyond the scope of the predefined keywords and uncover previously undocumented insights.
Using these keywords, we filtered 60,684 user reviews to 165, 749,067 Stack Overflow discussions to 2,158, 16,977 GitHub issues down to 149, and 344,922 GitHub commits to 558.

\subsubsection{Manual Checking.} \label{sec_manual_checking}
After filtering the data using performance-related keywords, we noticed some data entries were false positive, meaning they contained keywords but were not actually related to performance issues. 
Therefore, we further conducted a manual verification to ensure the data quality and relevance.

\noindent \textbf{User Reviews.} We removed user reviews that were unrelated to performance issues, such as criticisms of developers. For example, one review stated: \textit{``The app works well, but only supports SMBv1, which has been deprecated and known to be very insecure. There have been numerous complaints about this but the author has been stubbornly unresponsive about it...''}

\noindent \textbf{Question Threads.}
For Stack Overflow, we excluded question threads that are wrongly labeled on the \textit{code} tag as well as low-quality threads where the number of negative votes exceeded the number of positive votes.

\noindent \textbf{GitHub - Issues.} 
We first checked for and removed issues from inactive (dead) projects, such as those where developers do not respond or have explicitly stated that the project is no longer maintained.
Additionally, we excluded issues that do not involve performance issues, such as ``\textit{Show me ... ports mac address host name mDNS response time, perhaps option to ping again and option to traceroute.}''

\noindent \textbf{GitHub - Commits.} We manually checked all remaining commits and removed those that lacked clear messages. 
Additionally, we excluded commits with multiple changes, such as fixes for various bugs or features where performance was a minor aspect, as this made identifying the root cause difficult.

Ultimately, we obtained 114 user reviews, 1,484 Stack Overflow discussions, 69 GitHub issues, and 222 GitHub commits. 
Two of the authors performed the aforementioned manual checking individually, for any disagreement, they discussed their findings until reaching a consensus. For any persisting disagreement, a senior author made the final decision.
We utilized Cohen's kappa~\cite{Landis1977AnAO} to assess inter-rater agreement, yielding scores of 0.944 for user reviews, 0.868 for GitHub issues, 0.885 for GitHub commits, and 0.873 for Stack Overflow posts, all of which signify substantial agreement.
We then manually identified the Android performance issues within this data and present our findings in the following section.

\begin{equation*}
\begin{aligned}
    \text{Google Play - User Reviews: 60,684}  \rightarrow \text{165} &\rightarrow \text{114} \\
    \text{Stack Overflow - Question Threads: 749,067} \rightarrow \text{2,158} &\rightarrow \text{1,484} \\
    \text{GitHub - Issues \& Commits: 16,977/344,922}\rightarrow \text{149/558} &\rightarrow \text{69/222}
\end{aligned}
\end{equation*}

\subsection{Result Analysis} \label{sec:realworld data analysis}

To gain a thorough understanding of prevalent performance issues in real-world apps, we begin by analyzing the distribution of performance issues encountered by users and developers (Section~\ref{sec_pi_in_realworld}). 
Next, we categorize the underlying root causes contributing to these issues (Section~\ref{sec_realworld_root_causes}).
Finally, we identify code patterns commonly associated with these performance issues (Section~\ref{sec_common_pattern}).

\subsubsection{\textbf{Performance Issues in Real-world}} \label{sec_pi_in_realworld}

\begin{figure*}
    \centering
    \setlength{\belowcaptionskip}{-10pt}
    \includegraphics[width=0.98\textwidth]{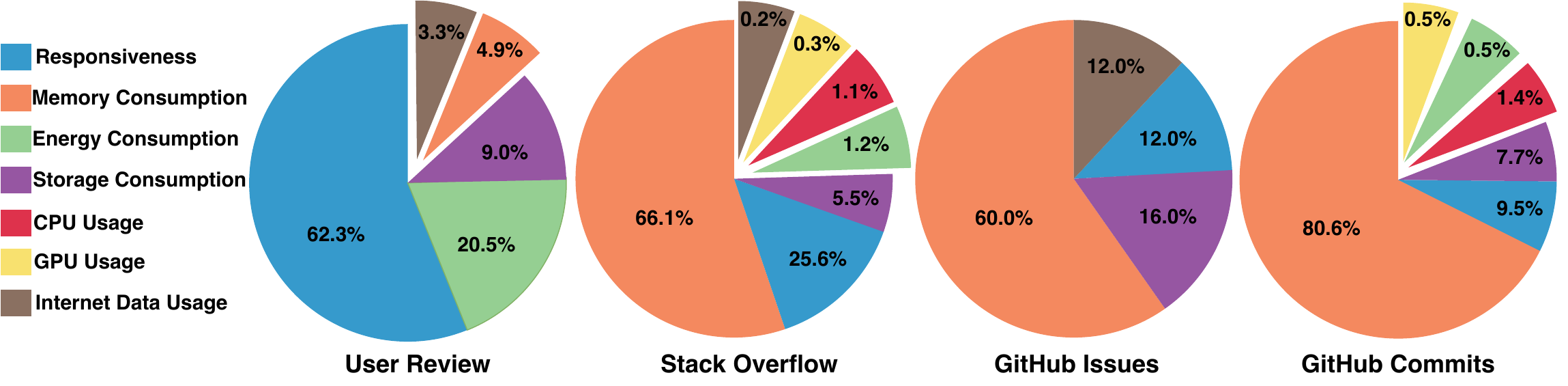}
    \caption{Distribution of performance issues across various sources.}

    \label{fig_realworld_pi_distribution}
\end{figure*}

We identified seven key performance issues that are of concern to users and frequently discussed by developers in real-world scenarios.
Fig.~\ref{fig_realworld_pi_distribution} shows the distribution of these issues across different platforms.

\noindent \textbf{[Google Play - User Reviews.]}
User reviews typically focus on how performance issues negatively impact their experience without delving into the specific causes. 
As shown in the \textit{User Review} pie chart in Fig.~\ref{fig_realworld_pi_distribution}, users are mainly concerned about \textit{responsiveness}, \textit{energy consumption}, \textit{storage consumption}, \textit{memory consumption}, and \textit{internet data usage}.
\textit{Responsiveness} issues, such as slow responsiveness or UI freezes, have the most direct impact on user experience, appearing in 62.3\% of all reported reviews.
This is not surprising as responsiveness is the most noticeable aspect of performance that users can directly perceive.
For example, one comment says:
\textit{``... the software appears to be sluggish in map updating and does not properly locate my vehicle... I am hopeful these issues will be resolved in the future but if not then I will stop discontinue my premium subscription service and find another solution.''}
This indicates that unresolved performance issues can lead to users canceling subscriptions, causing companies to lose market share.

\textit{Energy Consumption} is also critical, as inefficient app power consumption can quickly drain batteries. These issues appeared in 20.5\% of all reported reviews.
E.g., one representative comment states:
\textit{``severe battery drain on s10, causing 1\% loss every 3 minutes. drain immediately disappeared after uninstall, back to 1\% loss every 8-9 minutes.''}
This shows that performance issues can lead to users uninstalling the app.
\textit{Storage Consumption} is often complained by users that the size of the app is too large, or the cache space consumed is too large when using the app.
These issues appeared in 9.0\% of all reported reviews.
As mobile devices have limited storage, excessive storage consumption can lead to problems like a lack of available space for other apps or files, and potential app crashes.
E.g., one comment mentions:
\textit{``..Every time a full size version of a Google app gets updated it eats up even more internal storage space... if there's not enough storage space in your internal storage it won't install certain updates or new apps from Google Play store, you have to uninstall updates or apps before reinstalling updates and apps...''}
Finally, \textit{Internet Data Usage} issues are reported by users when apps consume excessive data, especially when they aren't using Wi-Fi, making them more sensitive to data usage.
For example, one user mentioned: 
\textit{``...Not sure why, but my data usage has skyrocketed. I'm on a limited plan, now I have to buy extra data every month...''}

\noindent \textbf{[Stack Overflow - Question Threads.]}
From the 1,484 question threads, we identified seven categories of performance issues, as illustrated in the \textit{Stack Overflow} pie chart in Fig.~\ref{fig_realworld_pi_distribution}.
Interestingly, two performance issues (i.e., \textit{CPU usage} and \textit{GPU usage}) are not mentioned in user reviews.
Additionally, while \textit{responsiveness} is the most concerning issue for end users, it's the second most frequent topic in developer discussions.
This discrepancy indicates that the performance issues developers focus on and those users are concerned about \textbf{do not align}.
This misalignment can be attributed to the differing perspectives of users and developers when evaluating performance.  
From a \textit{user-centric} perspective, performance is often assessed through perceptible outcomes such as smooth responsiveness, battery consumption, storage consumption, and internet data usage.  
In contrast, a \textit{developer-centric} perspective typically involves technical diagnostics, focusing on issues such as memory usage, CPU utilization, and GPU rendering efficiency.  
Since user satisfaction directly impacts app market share, developers should prioritize addressing the issues most important to users.

\noindent \textbf{[Github - Issues \& Commits.]}
The \textit{GitHub Issues} and \textit{GitHub Commits} pie charts in Fig.~\ref{fig_realworld_pi_distribution} show a similar distribution of performance issues across both data sources, with \textit{memory consumption}, \textit{responsiveness}, and \textit{storage consumption} being the most frequently mentioned.
Notably, while \textit{internet data usage} is a niche issue mainly mentioned in user reviews and absent from GitHub commits, it also appears in GitHub issues, which offer valuable insights into the underlying causes discussed by developers.

In summary, the five performance issues that users mention in reviews are also discussed by developers on GitHub and Stack Overflow. 
Developers further address two additional issues, \textit{CPU usage} and \textit{GPU usage}, which are rarely mentioned by users, likely because they are technical concerns not directly perceived in user experience.

\find{\textbf{Answers to RQ1.1}
\begin{itemize} [leftmargin=*]
    \item We identified seven categories of performance issues: \textit{responsiveness}, \textit{energy consumption}, \textit{storage consumption}, \textit{memory consumption}, \textit{internet data usage}, \textit{CPU usage}, and \textit{GPU usage}.
    \item A substantial divergence exists between the focus areas of performance issues for users and developers: while users primarily emphasize concerns around responsiveness, developers' discussions are predominantly oriented toward memory consumption.
\end{itemize}
}

\begin{table*}[]
\caption{Taxonomy of seven categories of performance issues and their contributing factors.}
\renewcommand{\arraystretch}{1.02}

\resizebox{\textwidth}{!}{
\small
\begin{tabular}{c|l|c|c|p{14.4cm}}
\hline
\rowcolor{white}
Category & Factor & Real-World & Literature & Problem Example \\ \hline

\multirow{22}{*}{\rotatebox[origin=c]{90}{\makecell[c]{Memory\\Consumption}}} 
     &  \cellcolor{gray!40} Loop & \cellcolor{gray!40} \ding{51} & \cellcolor{gray!40} \ding{55}   & \cellcolor{gray!40} Repeatedly reading the same value into the array in an infinite loop \\ \cline{2-5}
     &  \cellcolor{gray!40} Object Creation  &  \cellcolor{gray!40} \ding{51} & \cellcolor{gray!40} \ding{55}   & \cellcolor{gray!40}  Repeatedly creating new instances instead of reusing existing ones \\ \cline{2-5}
     & \cellcolor{gray!40} Massive Data &  \cellcolor{gray!40} \ding{51} & \cellcolor{gray!40} \ding{55}   & \cellcolor{gray!40} Outputting thumbnail URLs containing a large amount of information to the log \\\cline{2-5}
     &  \cellcolor{gray!40} Database Operation &  \cellcolor{gray!40} \ding{51} & \cellcolor{gray!40} \ding{55}   & \cellcolor{gray!40} Each database query loads new data into memory without clearing the previous data \\\cline{2-5}
     &  \cellcolor{gray!40} Recursive Call &  \cellcolor{gray!40} \ding{51} & \cellcolor{gray!40} \ding{55}   & \cellcolor{gray!40} Recursively sending requests rapidly creates excessive request queues and objects \\ \cline{2-5}     
     &  \cellcolor{gray!40} Activity &  \cellcolor{gray!40} \ding{51} & \cellcolor{gray!40} \ding{55}   & \cellcolor{gray!40} An anonymous inner class strongly references the Activity, preventing it from being garbage collected \\\cline{2-5}
     & \cellcolor{gray!40} File Handling &  \cellcolor{gray!40} \ding{51} & \cellcolor{gray!40} \ding{55}   & \cellcolor{gray!40} Loading an entire large file into memory at once, rather than processing it in chunks \\\cline{2-5}
     &  \cellcolor{gray!40} Fragment &  \cellcolor{gray!40} \ding{51} & \cellcolor{gray!40} \ding{55}   & \cellcolor{gray!40} The previous fragment state isn't properly cleared when switching between fragments \\\cline{2-5}
     &  \cellcolor{gray!40} Dynamic UI Loading &  \cellcolor{gray!40} \ding{51} & \cellcolor{gray!40} \ding{55}   & \cellcolor{gray!40} Recreating the view and bitmap for each list item during scrolling, without effectively reusing the views \\\cline{2-5}
     &  \cellcolor{gray!40} Background Thread  & \cellcolor{gray!40}  \ding{51} & \cellcolor{gray!40} \ding{55}   & \cellcolor{gray!40} Failing to stop the refresh when the app enters the background causes the UI component to remain active \\\cline{2-5}
     &  \cellcolor{gray!40} Singleton  & \cellcolor{gray!40}  \ding{51} & \cellcolor{gray!40} \ding{55}   & \cellcolor{gray!40} A singleton holds a reference to a UI component after its Activity is destroyed, preventing garbage collection \\\cline{2-5}

     &  \cellcolor{gray!40} Override  & \cellcolor{gray!40}  \ding{51} & \cellcolor{gray!40} \ding{55}   & \cellcolor{gray!40} Failing to call super.onDestroy() after overriding a method prevents proper release of system resources \\\cline{2-5}
     &  \cellcolor{gray!40} Configuration &  \cellcolor{gray!40} \ding{51} & \cellcolor{gray!40} \ding{55}   & \cellcolor{gray!40} Gradle JVM heap size is insufficient during project builds, causing OOM errors \\\cline{2-5}
     &  \cellcolor{gray!30} Image  & \cellcolor{gray!30} \ding{51} & \cellcolor{gray!30} \ding{51} &
     \cellcolor{gray!30} Decoding images without resizing; leaking image objects ~\cite{Song2021IMGDroidDI,Li2019CharacterizingAD,Carette2017InvestigatingTE,Kang2016ExperienceRD, Das2016AQA, MazueraRozo2020InvestigatingTA}\\\cline{2-5}
     &  \cellcolor{gray!30} Service  & \cellcolor{gray!30} \ding{51} & \cellcolor{gray!30} \ding{51} & 
     \cellcolor{gray!30}  Failing to destroy a dead service
     ~\cite{Song2019ServDroidDS}  \\\cline{2-5}
     & \cellcolor{gray!30}  Resource  & \cellcolor{gray!30} \ding{51}  & \cellcolor{gray!30} \ding{51} &
     \cellcolor{gray!30} Improper releasing of resources (e.g., Sensor, Camera) ~\cite{Liu2014GreenDroidAD,Liu2013WhereHM,Li2019DetectingAD,Behrouz2016EnergyawareTM,Li2022CombattingEI,Liu2016FixingRL,Couto2020EnergyRF,Jabbarvand2020AutomatedCO,Li2020EnergyDxDE,Lu2022DetectingRU,Banerjee2018EnergyPatchRR,Wu2016LightWeightIA,Jabbarvand2019SearchBasedET,Banerjee2014DetectingEB,Yan2013SystematicTF, Liu2019DroidLeaksAC, Das2016AQA, Wu2023ASL} \\\cline{2-5}
     &  \cellcolor{gray!30} AsyncTask & \cellcolor{gray!30} \ding{51}  & \cellcolor{gray!30} \ding{51} &
    \cellcolor{gray!30}  Failing to destroy AsyncTask when it holds active GUI components~\cite{Lin2015StudyAR,Pan2020StaticAC,Kang2016DiagDroidAP,Fan2018EfficientlyMA}\\\cline{2-5}
     & \cellcolor{gray!30}  View  & \cellcolor{gray!30} \ding{51} & \cellcolor{gray!30} \ding{51} &
     \cellcolor{gray!30} Making allocations inside onDraw() routines
     ~\cite{Habchi2019TheRO,Couto2020EnergyRF, Habchi2020AndroidCS} \\\cline{2-5}
    &  \cellcolor{gray!30} Third-Party Library  & \cellcolor{gray!30}  \ding{51} & \cellcolor{gray!30} \ding{51}   & \cellcolor{gray!30} The API usage in the new version of the third-party library caused excessive memory usage~\cite{Rua2023ALE} \\\cline{2-5}
    & \cellcolor{gray!30}  UI Rendering  & \cellcolor{gray!30}  \ding{51} & \cellcolor{gray!30} \ding{51}   & \cellcolor{gray!30} Reinflating views on each ``getView'' call without reusing them increases memory usage~\cite{Das2020CharacterizingTE} \\\cline{2-5}
    &  \cellcolor{gray!30} Data Type &  \cellcolor{gray!30} \ding{51} & \cellcolor{gray!30} \ding{51}   & \cellcolor{gray!30} Declaring static variables retains strong references to previous tasks, preventing garbage collection~\cite{Das2020CharacterizingTE, Palomba2017LightweightDO, Habchi2020AndroidCS} \\\cline{2-5}

    &  \cellcolor{gray!20} Obsolete Task & \cellcolor{gray!20} \ding{55} & \cellcolor{gray!20} \ding{51} &
    \cellcolor{gray!20} Retaining obsolete parameters after updating view layouts
     ~\cite{Kang2016DiagDroidAP,Couto2020EnergyRF} \\\hline

    \multirow{20}{*}{\rotatebox[origin=c]{90}{\makecell[c]{Responsiveness}}}

     & \cellcolor{gray!40}  Service & \cellcolor{gray!40} \ding{51} & \cellcolor{gray!40} \ding{55}  & \cellcolor{gray!40} Failing to move the notification to the foreground in time during service start, causing an ANR \\\cline{2-5}
     &  \cellcolor{gray!40} Inefficient Method & \cellcolor{gray!40} \ding{51} & \cellcolor{gray!40} \ding{55}  & \cellcolor{gray!40} Selecting a less efficient method introduces unnecessary delays in processing UI actions~\cite{Feng2019LearningPO} \\\cline{2-5}
     &  \cellcolor{gray!40} Background Thread &  \cellcolor{gray!40} \ding{51} & \cellcolor{gray!40} \ding{55}  & \cellcolor{gray!40} Rapidly creating numerous threads in the background to perform I/O on the same data \\\cline{2-5}

     &  \cellcolor{gray!40} Multi-Threads & \cellcolor{gray!40} \ding{51}  & \cellcolor{gray!40} \ding{55}  & \cellcolor{gray!40} A thread gets blocked while waiting for a shared mutex held by another thread \\\cline{2-5}

     &  \cellcolor{gray!40} Cache &  \cellcolor{gray!40} \ding{51} & \cellcolor{gray!40} \ding{55}  & \cellcolor{gray!40} Large cache causes frequent I/O operations, leading to lag \\\cline{2-5}
     & \cellcolor{gray!40}  Text Handling &  \cellcolor{gray!40} \ding{51} & \cellcolor{gray!40}  \ding{55}  & \cellcolor{gray!40} Repeatedly calling EditText.setText() to highlight all search results at once leads to unnecessary UI updates \\\cline{2-5}
     &  \cellcolor{gray!30} Resource &  \cellcolor{gray!30} \ding{51} & \cellcolor{gray!30} \ding{51}  & \cellcolor{gray!30} Insufficient thread resource allocation causes slow response during simultaneous operations~\cite{Bhatt2020AutomatedRO, Prestat2024DynAMICSAT} \\\cline{2-5}
     &  \cellcolor{gray!30} Image & \cellcolor{gray!30} \ding{51} & \cellcolor{gray!30} \ding{51} &
     \cellcolor{gray!30} Decoding large images in the UI thread~\cite{Song2021IMGDroidDI,Li2019CharacterizingAD,Carette2017InvestigatingTE,Kang2016ExperienceRD, Das2016AQA}\\\cline{2-5}
     & \cellcolor{gray!30}  Dynamic UI Loading & \cellcolor{gray!30} \ding{51} & \cellcolor{gray!30} \ding{51} &
      \cellcolor{gray!30} Frequent layout inflations during scrolling~\cite{Liu2014CharacterizingAD}  \\\cline{2-5}
     &  \cellcolor{gray!30} HTTP Request & \cellcolor{gray!30} \ding{51}  & \cellcolor{gray!30} \ding{51}  & \cellcolor{gray!30} Overlong connection timeout keeps page in loading state~\cite{Zhao2018LeveragingPA,Das2016AQA,Zhao2019SystematicallyTA, Das2016AQA} \\\cline{2-5}

     &  \cellcolor{gray!30} UI Rendering & \cellcolor{gray!30} \ding{51}  & \cellcolor{gray!30} \ding{51} &
     \cellcolor{gray!30} Improper layout stacking causes redundant pixel redraws in the same frame~\cite{Gao2017EveryPC,Lin2024AgingOG,Liu2014CharacterizingAD,Habchi2019TheRO, Vsquez2015HowDD, Rua2023ALE, Das2020CharacterizingTE, Janssen2022OnTI, Nistor2014SunCatHD, MazueraRozo2020InvestigatingTA, Wu2023ASL}\\\cline{2-5}
     &  \cellcolor{gray!30} Loop & \cellcolor{gray!30} \ding{51} & \cellcolor{gray!30} \ding{51}  & \cellcolor{gray!30} An infinite while loop in the button's onClick() event makes the app unresponsive after clicking~\cite{Lyu2018RemoveRF,Palomba2017LightweightDO, MazueraRozo2020InvestigatingTA} \\\cline{2-5}
     & \cellcolor{gray!30} Main Thread & \cellcolor{gray!30} \ding{51} & \cellcolor{gray!30} \ding{51} & \cellcolor{gray!30} Performing Heavy operations in the main thread ~\cite{Liu2014CharacterizingAD,Brocanelli2018HangDR,Lin2014RetrofittingCF, Gmez2016MiningTR, Vsquez2015HowDD, Das2020CharacterizingTE, Feng2019LearningPO, Das2016AQA, Wu2023ASL} \\\cline{2-5}
     &  \cellcolor{gray!30} Activity & \cellcolor{gray!30} \ding{51} & \cellcolor{gray!30} \ding{51} & \cellcolor{gray!30} Loading new resources during activity restarts may freeze the UI~\cite{Guo2022DetectingAF,Chen2023TransparentRC,Lu2022DetectingRU,Callan2022ImprovingRO} \\\cline{2-5}
     &  \cellcolor{gray!30} AsyncTask & \cellcolor{gray!30}  \ding{51} & \cellcolor{gray!30} \ding{51} &
     \cellcolor{gray!30} Processing tasks in a single-threaded pool of AsyncTask since Android 3.0~\cite{Lin2015StudyAR,Pan2020StaticAC,Kang2016DiagDroidAP,Fan2018EfficientlyMA, Prestat2024DynAMICSAT, Feng2019LearningPO, Das2016AQA}  \\\cline{2-5}
     &  \cellcolor{gray!30} View  & \cellcolor{gray!30} \ding{51} & \cellcolor{gray!30} \ding{51} & \cellcolor{gray!30} Allocating new objects along with draw operations~\cite{Habchi2019TheRO,Couto2020EnergyRF, Das2020CharacterizingTE, Prestat2024DynAMICSAT, Wu2023ASL}\\\cline{2-5}

     & \cellcolor{gray!20}  Obsolete Task & \cellcolor{gray!20} \ding{55} & \cellcolor{gray!20} \ding{51} &
    \cellcolor{gray!20} Downloading obsolete content occurs when switching activities
     ~\cite{Kang2016DiagDroidAP,Couto2020EnergyRF}\\\cline{2-5}

     & \cellcolor{gray!20}  SQL Statement & \cellcolor{gray!20}  \ding{55} & \cellcolor{gray!20} \ding{51}  &
     \cellcolor{gray!20} Retrieving more columns/rows than necessary in SQL queries
     ~\cite{Lyu2019QuantifyingTP}  \\\cline{2-5}

     &  \cellcolor{gray!20} Advertisements &  \cellcolor{gray!20} \ding{55} & \cellcolor{gray!20}  \ding{51}  &
     \cellcolor{gray!20} Loading ads while maintaining responsiveness to user actions~\cite{Petalotis2024AnES}  \\\cline{2-5}
     
     & \cellcolor{gray!20}  Message &  \cellcolor{gray!20}  \ding{55} & \cellcolor{gray!20} \ding{51}  &
     \cellcolor{gray!20}  Backloging rapidly incoming messages in Message Handler's single-threaded~\cite{Kang2016DiagDroidAP}\\\hline

\multirow{7}{*}{\rotatebox[origin=c]{90}{\makecell[c]{Storage\\Consumption}}}

    & \cellcolor{gray!40}  Cache & \cellcolor{gray!40} \ding{51} & \cellcolor{gray!40} \ding{55}  & \cellcolor{gray!40}  Residual data remains after the application is uninstalled  \\\cline{2-5}
    &  \cellcolor{gray!40} Third-Party Library & \cellcolor{gray!40} \ding{51} & \cellcolor{gray!40} \ding{55} & \cellcolor{gray!40} Unused third-party dependencies are packaged into the APK \\\cline{2-5}
    &  \cellcolor{gray!40} Resource & \cellcolor{gray!40} \ding{51} & \cellcolor{gray!40} \ding{55} & \cellcolor{gray!40} Unused resources (such as audio files and assets) are packaged into the APK \\\cline{2-5}
    & \cellcolor{gray!40}  File Handling & \cellcolor{gray!40} \ding{51} & \cellcolor{gray!40} \ding{55}  & \cellcolor{gray!40} Temp files from URIs aren't cleared after upload   \\\cline{2-5}
    & \cellcolor{gray!40}  Redundant Code & \cellcolor{gray!40} \ding{51} & \cellcolor{gray!40} \ding{55}  &  \cellcolor{gray!40} Similar functional pages and activities that could be reused, or unnecessary code that only occupies storage  \\\cline{2-5}
    &  \cellcolor{gray!40} Configuration & \cellcolor{gray!40} \ding{51} & \cellcolor{gray!40} \ding{55}  & \cellcolor{gray!40}  Including unnecessary third-party library dependencies in the Gradle file  \\\cline{2-5}
    & \cellcolor{gray!40}  Image & \cellcolor{gray!40} \ding{51} & \cellcolor{gray!40} \ding{55}  & \cellcolor{gray!40}  PNG images compress less effectively than JPG, resulting in larger file sizes  \\\hline

\multirow{20}{*}{\rotatebox[origin=c]{90}{\makecell[c]{Energy\\Consumption}}} 

     & \cellcolor{gray!30}  Resource & \cellcolor{gray!30} \ding{51}  & \cellcolor{gray!30} \ding{51} & 
     \cellcolor{gray!30} Updating unnecessary sensor data in the background ~\cite{Liu2014GreenDroidAD,Liu2013WhereHM,Behrouz2016EnergyawareTM,Li2022CombattingEI,Liu2016FixingRL,Couto2020EnergyRF,Jabbarvand2020AutomatedCO,Li2020EnergyDxDE,Guo2013CharacterizingAD,Lu2022DetectingRU,Banerjee2018EnergyPatchRR,Wu2016LightWeightIA,Jabbarvand2019SearchBasedET,Banerjee2014DetectingEB,Yan2013SystematicTF, Jabbarvand2017DroidAE, Nikzad2014APEAA, Wu2016Relda2AE, Vsquez2015HowDD, Bhatt2020AutomatedRO, Rua2023ALE, Liu2019DroidLeaksAC, Hort2021ASO, LeGoaer2022ecoCodeAS, Cruz2019CatalogOE, BernalCrdenas2015ImprovingEC, Goar2020EnforcingGC, Vsquez2017GEMMAMO, Morales2018EARMOAE, Das2016AQA} \\\cline{2-5}
     & \cellcolor{gray!30}  Background Thread & \cellcolor{gray!30} \ding{51} & \cellcolor{gray!30} \ding{51}  &   
     \cellcolor{gray!30} Setting background services at unnecessarily high frequencies~\cite{Liu2014CharacterizingAD,Banerjee2018EnergyPatchRR,Li2019DetectingAD,Banerjee2014DetectingEB,Jabbarvand2019SearchBasedET, BernalCrdenas2015ImprovingEC} \\\cline{2-5}
     & \cellcolor{gray!30}  Wakelock & \cellcolor{gray!30}  \ding{51}  & \cellcolor{gray!30} \ding{51}  & \cellcolor{gray!30}  Forgetting to release wakelocks after usage ~\cite{Liu2014GreenDroidAD,Pathak2012WhatIK,Liu2016UnderstandingAD,Li2022CombattingEI,Couto2020EnergyRF,Li2020EnergyDxDE,Behrouz2016EnergyawareTM,Banerjee2018EnergyPatchRR,Banerjee2014DetectingEB,Pathak2012WhereIT, Jabbarvand2017DroidAE, Vsquez2015HowDD, Das2020CharacterizingTE, LeGoaer2022ecoCodeAS, Cruz2019CatalogOE, Goar2020EnforcingGC, Iannone2020RefactoringAE, Vsquez2017GEMMAMO, Prestat2024DynAMICSAT, Palomba2017LightweightDO, MazueraRozo2020InvestigatingTA, Wu2023ASL} \\\cline{2-5}

     & \cellcolor{gray!30}  HTTP Request & \cellcolor{gray!30} \ding{51}  & \cellcolor{gray!30} \ding{51}  &    
     \cellcolor{gray!30} Dividing a large packet into small pieces and transmitting data frequently~\cite{Li2016AutomatedEO, Li2014AnES, Goar2020EnforcingGC, Vsquez2017GEMMAMO, Das2016AQA, Wu2023ASL}
     \\\cline{2-5}
     & \cellcolor{gray!30}  Loop &  \cellcolor{gray!30} \ding{51}  & \cellcolor{gray!30}  \ding{51}  & 
     \cellcolor{gray!30} Executing energy-intensive behaviors in loop ~\cite{Banerjee2014DetectingEB,Behrouz2016EnergyawareTM,Lyu2018RemoveRF,Fratantonio2015CLAPPCL,Lyu2017AnES, Jabbarvand2017DroidAE, Li2014AnES, Palomba2019OnTI, Goar2020EnforcingGC, Das2016AQA}\\\cline{2-5}
     & \cellcolor{gray!30}  Service &   \cellcolor{gray!30} \ding{51}  & \cellcolor{gray!30} \ding{51}  & 
     \cellcolor{gray!30} Creating or stopping services at an inappropriate timeslot ~\cite{Song2019ServDroidDS, BernalCrdenas2015ImprovingEC, Goar2020EnforcingGC, Das2016AQA}\\\cline{2-5}
     
     & \cellcolor{gray!20}  Redundant Frames & \cellcolor{gray!20} \ding{55}  & \cellcolor{gray!20} \ding{51}  & \cellcolor{gray!20}   Generating frames even when there are no visual changes ~\cite{Li2019DetectingAD,Kim2016ContentCentricEM,LeGoaer2022ecoCodeAS}\\\cline{2-5}
     & \cellcolor{gray!20}  Database Operation & \cellcolor{gray!20}  \ding{55}  & \cellcolor{gray!20} \ding{51}  & \cellcolor{gray!20} Frequently triggering database queries for auto-complete~\cite{Li2019DetectingAD,Lyu2018RemoveRF,Li2022CombattingEI,Lyu2017AnES}  \\\cline{2-5}
     & \cellcolor{gray!20}  Multi-Threads & \cellcolor{gray!20}  \ding{55}  & \cellcolor{gray!20} \ding{51}  &  \cellcolor{gray!20} Activating/deactivating a UI component simultaneously in two different threads~\cite{Pathak2012WhatIK, Feng2019LearningPO, Das2016AQA}  \\\cline{2-5}
     &  \cellcolor{gray!20} Inefficient Method & \cellcolor{gray!20}  \ding{55}  & \cellcolor{gray!20} \ding{51}  & \cellcolor{gray!20} Calling unnecessary methods~\cite{Carette2017InvestigatingTE,Couto2020EnergyRF,Habchi2019TheRO, Palomba2019OnTI, Vsquez2014MiningEA, Goar2020EnforcingGC, Iannone2020RefactoringAE, Vsquez2017GEMMAMO, Bangash2023CosteffectiveSF, Prestat2024DynAMICSAT, Morales2018EARMOAE, Palomba2017LightweightDO, Feng2019LearningPO, Das2016AQA, Wu2023ASL} \\\cline{2-5}

     & \cellcolor{gray!20}  Obsolete Task &  \cellcolor{gray!20}  \ding{55}  & \cellcolor{gray!20} \ding{51}  & 
     \cellcolor{gray!20} Downloading unnecessary content when switching activities
     ~\cite{Kang2016DiagDroidAP,Couto2020EnergyRF,Cruz2019CatalogOE} \\\cline{2-5}
     & \cellcolor{gray!20}  Image &  \cellcolor{gray!20}  \ding{55}  & \cellcolor{gray!20} \ding{51}  & 
     \cellcolor{gray!20} Decoding identical image in multiple times~\cite{Song2021IMGDroidDI,Li2019CharacterizingAD,Carette2017InvestigatingTE,Kang2016ExperienceRD, Das2016AQA} \\\cline{2-5}

     & \cellcolor{gray!20}  Dynamic UI Loading &  \cellcolor{gray!20} \ding{55}  & \cellcolor{gray!20} \ding{51}  & 
     \cellcolor{gray!20} Frequent layout inflations during scrolling
     ~\cite{Liu2014CharacterizingAD} \\\cline{2-5}
     & \cellcolor{gray!20}  UI Rendering &  \cellcolor{gray!20} \ding{55}  & \cellcolor{gray!20} \ding{51}  &  \cellcolor{gray!20} Loading CSS for below-the-fold content during webpage loading wastes energy
     ~\cite{Gao2017EveryPC,Lin2024AgingOG,Liu2014CharacterizingAD,Habchi2019TheRO, Janssen2022OnTI, Cruz2019CatalogOE, Wu2023ASL} \\\cline{2-5}
     &  \cellcolor{gray!20} AsyncTask&   \cellcolor{gray!20} \ding{55}  & \cellcolor{gray!20} \ding{51}  & 
    \cellcolor{gray!20} Updating destroyed components during AsyncTask life-cycle~\cite{Lin2015StudyAR,Kang2016DiagDroidAP,Pan2020StaticAC,Lin2014RetrofittingCF,Fan2018EfficientlyMA, BernalCrdenas2015ImprovingEC, Iannone2020RefactoringAE}  \\\cline{2-5}

    & \cellcolor{gray!20}  Display Screen & \cellcolor{gray!20}  \ding{55}  & \cellcolor{gray!20}  \ding{51}  & \cellcolor{gray!20}  The OLED display consumes more energy when displaying light colors than dark ones~\cite{Li2014MakingWA, LinaresVsquez2018MultiObjectiveOO, Vsquez2015OptimizingEC, Wan2015DetectingDE, Cruz2019CatalogOE, BernalCrdenas2015ImprovingEC, Goar2020EnforcingGC, Vsquez2017GEMMAMO} \\\cline{2-5}
    & \cellcolor{gray!20}  Program Language &  \cellcolor{gray!20} \ding{55}  & \cellcolor{gray!20} \ding{51}  & \cellcolor{gray!20}  Developing web applications using JavaScript consumes more energy than WebAssembly~\cite{Hasselt2022ComparingTE, Oliveira2017ASO} \\\cline{2-5}
    & \cellcolor{gray!20}  Advertisements & \cellcolor{gray!20}  \ding{55}  & \cellcolor{gray!20} \ding{51}  &  \cellcolor{gray!20} Frequent advertisement and analytics (A\&A) service requests lead to energy waste~\cite{Cito2016BatteryawareTI, Petalotis2024AnES, Vsquez2017GEMMAMO} \\\cline{2-5}
    & \cellcolor{gray!20}  Machine Learning & \cellcolor{gray!20}  \ding{55}  & \cellcolor{gray!20} \ding{51}  &  \cellcolor{gray!20} The inefficient machine learning algorithm on mobile devices is overly complex~\cite{McIntosh2018WhatCA} \\\cline{2-5}
    & \cellcolor{gray!20}  View &  \cellcolor{gray!20}  \ding{55}  & \cellcolor{gray!20} \ding{51}  & 
     \cellcolor{gray!20} Drawing all items in a ListView without reusing exising data
     ~\cite{Habchi2019TheRO,Couto2020EnergyRF, Feng2019LearningPO}  \\\hline

\multirow{7}{*}{\rotatebox[origin=c]{90}{\makecell[c]{CPU\\Usage}}} 

    & \cellcolor{gray!40}  Multi-Threads & \cellcolor{gray!40} \ding{51} & \cellcolor{gray!40} \ding{55}  & \cellcolor{gray!40}  The music player mishandles metadata extraction, spawning thousands of threads, overloading the CPU  \\\cline{2-5}
    &  \cellcolor{gray!40} Loop & \cellcolor{gray!40} \ding{51} & \cellcolor{gray!40} \ding{55}  &  \cellcolor{gray!40} An infinite loop repeatedly calls methods to handle images  \\\cline{2-5}
    & \cellcolor{gray!40}  Image & \cellcolor{gray!40} \ding{51} & \cellcolor{gray!40} \ding{55}  &  \cellcolor{gray!40} Glide consumes significant CPU resources to sequentially display images when playing GIFs  \\\cline{2-5}
    &  \cellcolor{gray!40} Third-Party Library  & \cellcolor{gray!40}  \ding{51} &  \cellcolor{gray!40} \ding{55} &  \cellcolor{gray!40} Third-party library’s bug causes  a failure to properly release the resources used to display GIFs  \\\cline{2-5}
    & \cellcolor{gray!40}  Background Thread  & \cellcolor{gray!40} \ding{51}  & \cellcolor{gray!40} \ding{55}  &  \cellcolor{gray!40} Frequently perform read and write operations in a background thread  \\\cline{2-5}
    & \cellcolor{gray!40}  Resource  & \cellcolor{gray!40} \ding{51}  & \cellcolor{gray!40}  \ding{55} & \cellcolor{gray!40} Use high sampling rate sensors such as the gyroscope  \\\cline{2-5}
    & \cellcolor{gray!30}  UI Rendering & \cellcolor{gray!30} \ding{51} & \cellcolor{gray!30} \ding{51}  & \cellcolor{gray!30} Unnecessarily using transparency to render UI components~\cite{Lee2019ImprovingEE}    \\\hline

\multirow{2}{*}{\rotatebox[origin=c]{90}{\makecell[c]{GPU\\\scalebox{0.9}{Usage}}}} 

    & \cellcolor{gray!40}  Third-Party Library & \cellcolor{gray!40} \ding{51} & \cellcolor{gray!40} \ding{55}  & \cellcolor{gray!40} Third-party library's bug causes continuous redrawing when using GPU to render UI components \\\cline{2-5}

    & \cellcolor{gray!30}  UI Rendering & \cellcolor{gray!30} \ding{51} & \cellcolor{gray!30} \ding{51}  & \cellcolor{gray!30} Spin animation has unnecessary frequent refreshes, forcing GPU rendering on every frame~\cite{Lee2019ImprovingEE, Habchi2020AndroidCS, Wu2023ASL}  \\\hline

\multirow{4}{*}{\rotatebox[origin=c]{90}{\makecell[c]{Internet\\Data\\Usage}}} 
    & \cellcolor{gray!40}  Resource & \cellcolor{gray!40} \ding{51} & \cellcolor{gray!40} \ding{55}  & \cellcolor{gray!40} Downloading unwanted content without user permission, wasting internet resources \\\cline{2-5}
    & \cellcolor{gray!40}  Image & \cellcolor{gray!40} \ding{51} & \cellcolor{gray!40} \ding{55}  &  \cellcolor{gray!40} Automatically loading animated GIFs without Wi-Fi enabled  \\\cline{2-5}
    & \cellcolor{gray!40}  HTTP Request & \cellcolor{gray!40} \ding{51} & \cellcolor{gray!40} \ding{55}  & \cellcolor{gray!40} High-speed movement triggers frequent GPS updates, causing redundant API uploads  \\\cline{2-5}

    & \cellcolor{gray!40}  Advertisements & \cellcolor{gray!40} \ding{51} & \cellcolor{gray!40} \ding{55}  & \cellcolor{gray!40} Automatically downloading embedded advertising services in the background  \\\hline


\end{tabular}
}

\label{tab:performance__issues}

\end{table*}

\subsubsection{\textbf{Root Causes of Performance Issues}} \label{sec_realworld_root_causes}
To address performance issues, it is crucial to investigate the underlying root causes that lead to these issues.
Based on our real-world analysis, we manually summarized the root causes (i.e., contributing factors) for each performance category, as indicated by the checkmarks (\ding{51}) in the \textit{R-W.} column in Table~\ref{tab:performance__issues}.

Here, we remind readers that during the process of manual summarization, we found that the descriptions of performance issues are chaotic.
The absence of a universally accepted taxonomy results in many discussions targeting the same performance problems but using different descriptive terms~\cite{Hort2021ASO, Liu2014CharacterizingAD, Habchi2019TheRO, Rua2023ALE, Das2020CharacterizingTE}. Additionally, since one factor can lead to multiple performance issue consequences, and one consequence can be caused by multiple factors, it complicates the creation of a clear and comprehensible performance classification system.
To address this gap, our work distinguishes between the consequences and factors of performance issues.

Specifically, we applied the thematic analysis methodology~\cite{Cruzes2011RecommendedSF, Das2016AQA} to analyze all the real-world data we collected, including GitHub issues and commits, Google Play Store user comments, and Stack Overflow posts. 
Using this approach, we classified performance issues according to their observed consequences, identified the underlying contributing factors, and established explicit links between them to construct a unified cross-source taxonomy.

This process involved the following steps:
\begin{itemize} [leftmargin=*]
    \item (i) \textbf{Familiarization with the data:} 
    Two authors independently reviewed all the real-world records to gain a thorough understanding of the content.
    \item (ii) \textbf{Initial coding:} 
    After familiarization, the two authors conducted open labeling by assigning descriptive tags to each record.
    In total, they analyzed 114 Google Play user reviews, 1,484 Stack Overflow question threads, 69 GitHub issues, and 222 GitHub commits, as described in Section~\ref{sec_manual_checking}.
    Each fragment was labeled independently by both authors. 
    At this stage, we identified 226 distinct labels. 
    For example, a review, GitHub issue, Stack Overflow post, or commit might be annotated with tags such as \textit{memory usage inefficiency}, \textit{memory leak}, \textit{out of memory}, \textit{image decoding without resizing}, or \textit{image decoding in the main thread}.
    \item (iii) \textbf{Categorization of issues and factors:}
    The initial labels were then organized into higher-level performance issues and contributing factors. 
    For instance, \textit{memory usage inefficiency}, \textit{memory leak}, and \textit{out of memory} were grouped under the broader issue category \textit{Memory Consumption}. 
    In contrast, tags such as \textit{decoding images without resizing}, \textit{inefficient image format}, and \textit{decoding identical images multiple times} were consolidated into the broader \textit{Image} factor. 
    At this stage, we obtained 15 categories of performance issues and 96 categories of contributing factors.
    \item (iv) \textbf{Review and refinement:} 
    All authors collaboratively reviewed and refined the labeled data to ensure accurate classification.
    During this iterative process, we held multiple discussions to merge semantically similar items, split overly broad categories, and discard irrelevant or redundant tags. 
    For example, we combined the factors \textit{Drawing View} and \textit{Bottom Sheet View} into the broader \textit{View} category.
    Finally, through this systematic and iterative process, we developed a taxonomy of seven major performance issues and 63 contributing factors.
    An illustrative example is presented in Table~\ref{tab:performance__issues}, where the \textit{Real-World} column is marked with (\ding{51}).
\end{itemize}

\textbf{Memory Consumption.}
As discussed in Section~\ref{sec_pi_in_realworld}, memory consumption is the most frequently highlighted performance issue among developers.
Our real-world analysis revealed 21 factors that contribute to these memory consumption issues, which are listed under the \textit{Memory Consumption} category in Table~\ref{tab:performance__issues}.
These issues arise from a variety of factors, often resulting from inefficient coding practices that lead to excessive or unnecessary memory use. 
For example, repeatedly reading the same value into an array within an infinite loop can cause memory to be continuously occupied without being released, ultimately leading to an out-of-memory (OOM) error. 
Additionally, failing to destroy an AsyncTask when it holds references to active GUI components can prevent the garbage collection of the destroyed GUI, leading to memory leaks.

\textbf{Responsiveness.}
Another important aspect of Android performance issues is responsiveness. 
Poor responsiveness leads to perceptible delays such as image rendering delays, UI freezes, unresponsiveness, etc.
In our analysis, we identified 16 key factors contributing to this issue.
Many real-world scenarios show that heavy operations associated with UI threads can slow responsiveness, such as decoding large images in the UI threads.
Besides, inefficient code logic can also slow down the response time.
For example, improper stacking and layering in layouts can cause pixels to be redrawn multiple times within the same frame, slowing down the response time.

\textbf{Storage Consumption.}
As the third most concerning performance issue for developers, inefficient storage consumption often leads to apps occupying more storage space than expected, which can limit the device's ability to install other apps and store data.
Through our exploration, we identified seven key factors contributing to storage consumption issues.
These factors typically arise from inefficient coding practices and poor data management, resulting in excessive or unnecessary storage usage. 
For instance, creating a large number of similar functional pages and activities that could have been reused, or including redundant code that never gets executed during runtime, wastes storage space.
Moreover, adding unnecessary third-party library dependencies to the Gradle file further increases the app's size after building, leading to larger APK files.

\textbf{Energy Consumption.}
As outlined in Section~\ref{sec_pi_in_realworld}, energy consumption ranks as the second highest concern among users, as apps that drain battery faster than expected can frustrate users, especially given the limited battery resources on mobile devices.
However, developers have not prioritized addressing this issue to the same extent.
Consequently, in our real-world analysis, we identified only six key factors contributing to energy consumption problems.
The root causes generally stem from inefficient resource usage and unnecessary background operations that provide no tangible benefits to users. 
For instance, updating unneeded sensor data in the background or running background services at unnecessarily high frequencies both result in excessive energy consumption.
It is worth noting that, although developers do not prioritize energy efficiency, many research papers have extensively studied this issue, bridging the gap in developer attention. 
We will further discuss the performance issues targeted by researchers in Section~\ref{sec:problem}.

\textbf{CPU Usage.}
Excessive CPU usage can cause the device to overheat, triggering the thermal management system to automatically lower the processor's frequency to prevent overheating.
This leads to a decline in overall system performance, affecting the normal operation of other apps and background services, and may even result in crashes, while also shortening the lifespan of the hardware and battery.
Based on our analysis, we identified seven distinct factors contributing to CPU usage issues.
These issues typically stem from unnecessary or inefficient operations that waste CPU resources. 
For instance, using transparency in canvases where it’s not needed forces the browser to perform extra work, increasing CPU load. 
Similarly, auto-playing GIFs by sequentially displaying images instead of using a static preview image results in unnecessary CPU consumption.

\textbf{GPU Usage.}
The GPU is essential for handling graphics-intensive tasks like UI rendering, animations, 3D graphics, and video/image processing, reducing the CPU’s workload. 
Improper or excessive GPU use can lead to issues like UI freeze, frame drops, device overheating, rapid battery drain, and inefficient system resource allocation.
Our real-world analysis identified two key factors contributing to GPU usage issues.
These issues typically arise from overly frequent rendering requests or third-party libraries' bugs.
For example, a spin animation with excessively frequent refreshes forces the GPU to render every frame, unnecessarily consuming resources. 
Besides, a bug in a third-party library could cause an app's misbehavior such as continuous redrawing of UI components, resulting in excessive GPU usage without providing any tangible benefits.

\textbf{Internet Data Usage.}
This category primarily refers to the unnecessary or unauthorized use of network resources, regardless of whether the device is connected via Wi-Fi or cellular data, and regardless of whether the network behavior is caused by the app's own logic or by integrated third-party APIs.
Although such issues may occur under various network conditions, their negative impact is often more significant when users are not connected to Wi-Fi, as inefficient data usage can lead to unexpected expenses.
From our analysis, we identified four key factors contributing to this issue.
The root causes often involve unnecessary or unauthorized consumption of network resources, regardless of whether the device is connected via Wi-Fi or cellular data, and irrespective of whether the behavior is initiated by the app itself or by third-party libraries (e.g., advertising SDKs).
For example, downloading unwanted content without user consent or automatically downloading embedded advertisements in the background can lead to excessive data usage.

\find{\textbf{Answers to RQ1.2:}
\begin{itemize} [leftmargin=*]
    \item Performance issues are inherently complex, often arising from an interplay of multiple contributing factors, where a single issue may stem from various causes, and, conversely, a single factor may amplify multiple issues.
    \item  We identified 63 distinct factors contributing to the seven categories of performance issues.
\end{itemize}
}

\subsubsection{\textbf{Common Code Patterns in Real-World}} \label{sec_common_pattern}

After identifying the key contributing factors of performance issues, we analyzed Stack Overflow discussions, GitHub issues, and commits to uncover common code patterns associated with these issues.
We found that the relationships among \textit{code patterns}, \textit{performance issues}, and \textit{contributing factors} are inherently \textbf{many-to-many}:
a single code pattern may involve multiple contributing factors and lead to various types of performance issues; similarly, a single factor may appear across multiple different code patterns.
It is worth noting that a number of prior studies have also focused on identifying and summarizing performance-related code patterns in Android applications~\cite{Das2020CharacterizingTE, cruz2018using, Liu2014CharacterizingAD, Nistor2014SunCatHD, Guo2013CharacterizingAD, Song2021IMGDroidDI, Habchi2019TheRO, Li2019CharacterizingAD, Liu2016UnderstandingAD, Cruz2019CatalogOE}.
For example, the well-known \textit{DrawAllocation} issue typically refers to performing memory allocation within UI rendering methods such as \texttt{onDraw()}, and has been widely discussed in the literature~\cite{Das2020CharacterizingTE, cruz2018using} and flagged by the Android Lint tool.
This specific issue often involves contributing factors such as \textit{Loop}, \textit{Object Creation}, and \textit{UI Rendering}.
In our study, however, \textit{DrawAllocation} is considered a concrete instance of a broader code pattern category we term \textit{UI Operations}.
More generally, each of our identified code pattern categories encompasses a broader set of real-world cases, which may be composed of different combinations of contributing factors.
Many specific patterns reported in prior research can be viewed as subtypes under these broader categories.
We present six representative categories of common code patterns below.

\begin{wrapfigure}{r}{0.5\textwidth} 
\vspace{-10pt}
\begin{lstlisting}[language=Typescript, escapechar=\%, caption=Code Example of API Order Misuse, label=lst:copytext, numbers=right]
const CopyTextOption = ({bottomSheetId, postMessage, sourceScreen}: Props) => {
    const handleCopyText = useCallback(async () => {
        ...
        await dismissBottomSheet(bottomSheetId);
        Clipboard.setString(postMessage);    
    };};  
\end{lstlisting}
\vspace{-8mm}
\end{wrapfigure}

\textbf{API Misuse.}
API misuse presents a significant concern in relation to performance issues, encompassing misuses such as incorrect API invocation, misuse of API orders, and misuse of API parameters. These errors can result in severe performance consequences, such as ANR and delays. 
For example, passing a null context (a bad practice in parameter value passing) to \emph{PendingIntent} may result in an Application Not Responding (ANR) error. 
This pattern may involve multiple contributing factors. 
Specifically, if the context is derived from an \textit{Activity} or \textit{Service} with an invalid lifecycle state, it can prevent the intent from being properly dispatched or executed, thereby blocking the \textit{Main Thread} and eventually triggering an ANR.
Besides, another common pattern involves the misuse of API order, as demonstrated in the code snippet~\ref{lst:copytext} from an
app \emph{com.mattermost.rnbeta}, where \textit{dismissBottomSheet} (line 4) is invoked before \textit{Clipboard.setString(postMessage)} (line 5). In this scenario, \textit{Clipboard.setString} has to wait until \textit{dismissBottomSheet} is executed, leading to a delay in response time.
This case mainly involves the \textit{Main Thread} and \textit{UI Rendering} factors.

\begin{wrapfigure}{r}{0.5\textwidth} 
\vspace{-10pt}
\begin{lstlisting}[language=Java, caption= Code Example of Unreleased Reference, label=lst:context reference, numbers=right, numbersep=-2pt]
public class MySingleton() {
    Private Context mContext;
    Private static MySingleton mInstance;
    ...
    public void myMethod(Context context)
    {this.mContext = context;...}}
\end{lstlisting}

\vspace{-6mm}

\end{wrapfigure}

\textbf{Unreleased References.}
Another significant pattern can be characterized by the failure to release references (or have them garbage collected). For example, if a Fragment is held as an implicit reference, it won't be garbage collected until the object holding the reference completes its execution. Initializing objects in complex Activity lifecycle methods can also result in unreleased references, ending up with memory leak. Listing~\ref{lst:context reference} demonstrates a \href{https://stackoverflow.com/questions/56102382}{code snippet} from Stack Overflow,
where a singleton object holds an \emph{mContext} reference (line 2). Here, \emph{mContext} is a layer that stands behind its component (Activity, Service, etc.), and its lifecycle is much shorter than that of the singleton object. This prevents it from being garbage collected, leading to a memory leak.
This case mainly involves the \textit{Singleton}, \textit{Activity}, and \textit{Service} factors.

\begin{wrapfigure}{r}{0.5\textwidth} 
\vspace{-10pt}
\begin{lstlisting}[language=Java, caption=Code Example of Redundant Objects, escapechar=\%, label=lst:textview, numbers=right, numbersep=-2pt]
public class MyTextView extends TextView {
    public void setTypeface(Typeface tf, int style) {
        if (style == Typeface.BOLD) {
            super.setTypeface(%\underline{Typeface.createFromAsset(...)}%);
        } else if (style == Typeface.ITALIC) {
            super.setTypeface(%\underline{Typeface.createFromAsset(...)}%);
        } else {
            super.setTypeface(%\underline{Typeface.createFromAsset(...)}%);}}}
\end{lstlisting}

\vspace{-6mm}

\end{wrapfigure}

\textbf{Redundant Objects.}
The creation of redundant objects is a common pattern that causes performance issues. Developers often mistakenly create new instances repeatedly instead of referring to existing ones, leading to endless memory consumption. Additionally, when two classes are mutually dependent and create instances of each other during initialization, it results in endless object creation. Moreover, creating new objects with the same functionality can also unnecessarily degrade performance.
Listing~\ref{lst:textview} demonstrates a \href{https://stackoverflow.com/questions/20942671}{code snippet} from Stack Overflow,
where every time \textit{setTypeface} (line 2) is called, it creates a new \textit{Typeface} object (line 4, 6, 8). This leads to increased memory usage, app slowdowns, and eventual out-of-memory crashes.
This case mainly involves the \textit{Object Creation} factor.

\begin{wrapfigure}{r}{0.5\textwidth} 
\vspace{-2pt}
\begin{lstlisting}[language=Java, escapechar=\%, caption=Code Example of Large-Scale Data, label=lst:largesize file, numbers=right, numbersep=-2pt]
    ...
    int length = 0;
    newFile.createNewFile();
    InputStream inputStream = ctx.getAssets().open("myBigFile.apk");
    FileOutputStream fOutputStream = new FileOutputStream(newFile);
    byte[] buffer = new byte[%\underline{inputStream.available()}%];
    while ((length = inputStream.read(buffer)) > 0) {
        fOutputStream.write(buffer, 0, length);}
\end{lstlisting}

\vspace{-6mm}
\end{wrapfigure}

\textbf{Large-Scale Data.}
Large-scale data is also a common cause of performance issues, since operations on such data often exceed the capacity of mobile devices with limited resources. 
Loading large images without proper caching can cause excessive memory usage and slowdowns. Uploading large files without separating them into smaller chunks can lead to high memory consumption. As shown in a \href{https://stackoverflow.com/questions/8820837}{code snippet} from Stack Overflow,
where the buffer size is the same as the uploaded file. If the uploaded file exceeds the maximum available memory, it will directly cause an out-of-memory error.
This case mainly involves the \textit{Massive Data} and \textit{File Handling} factors.

\begin{wrapfigure}{r}{0.5\textwidth} 
\vspace{-10pt}
\begin{lstlisting}[language=Java, caption=Drawing the Same Content in the Loop, label=lst:draw same content, numbers=right, numbersep=-2pt]
public class ConcentricCircularView extends View {...
  protected void onDraw(Canvas canvas) {
    super.onDraw(canvas);
    ...
    for (int i=0;i<10;i++){...}}} 
\end{lstlisting}

\vspace{-6mm}

\end{wrapfigure}
\textbf{UI Operations.}
UI is another key factor that can affect performance, encompassing issues such as repeatedly drawing the same content and UI layout misconfiguration. For example, in the \href{https://stackoverflow.com/questions/42944376}{code snippet}~\ref{lst:draw same content} from Stack Overflow,
the `for' loop in `onDraw' (line 5) redraws the same circles every time when \emph{onDraw} is invoked, even though the UI content doesn't change. This causes the application to slow down and increases the computational load.
This case mainly involves contributing factors such as \textit{Loop} and \textit{UI Rendering}.

\textbf{Other Patterns.}
Besides the previously mentioned patterns, other types of patterns also warrant attention, such as scheduling a large number of \emph{Runnables} to execute simultaneously, accessing \emph{CookieManager} on the main thread, etc. 
We made the full list publicly available in our artifact:~\cite{our_repo}.

\find{\textbf{Answers to RQ1.3}
We identified six prevalent code patterns that commonly contribute to performance issues: \textit{API Misuse}, \textit{Unreleased References}, \textit{Redundant Objects}, \textit{Large-Scale Data}, \textit{UI Operations}, and \textit{Other Patterns}.
}

\subsection{Practitioner Validation of the Taxonomy} \label{sec:dev_validation}

To further validate whether the performance issues and contributing factors identified in our taxonomy reflect practical concerns, we conducted a survey with Android practitioners.

\subsubsection{Method}
We designed a questionnaire consisting of the following parts:
\begin{itemize}[leftmargin=*]
\item \textbf{Q1a:} What is your current role related to Android software development?
\item \textbf{Q1b:} What is your age range? (23 or below, 24–27, 28 or above)
\item \textbf{Q1c:} How many years of experience do you have in Android development or related research?
\item \textbf{Q2:} This section presents seven common categories of performance issues. For each category, please indicate the extent to which it concerns you during development. Responses were recorded on a 5-point Likert scale: 1 = not concerned at all, 2 = slightly concerned, 3 = moderately concerned, 4 = concerned, 5 = very concerned.  
\item \textbf{Q3:} This section presents each performance issue category along with its contributing factors and an example illustrating how each factor can lead to the issue.
For each category, you are asked to review the associated factors and select all those that they consider a concern or believe may potentially affect performance during development.
This question used a multiple-choice format, allowing participants to select zero or more factors that reflect their concerns or are potentially relevant to their development experience.  
\end{itemize}
Among them, Q2 was designed to validate the relevance of the seven performance issue categories by measuring how strongly participants with Android development experience perceive each category as a practical concern. 
Q3 was designed to validate the comprehensiveness of the contributing factors by examining whether participants recognize these factors as relevant to their own development experience, thereby assessing the alignment between our taxonomy and real-world practice.
The survey was conducted online and remained open for two weeks. 
We collected anonymous responses to encourage honest feedback, and all participants provided informed consent before answering the questions.

\begin{wrapfigure}{r}{0.5\textwidth}
    \centering
    \setlength{\abovecaptionskip}{-2pt}
    \setlength{\belowcaptionskip}{-10pt}
    \includegraphics[width=0.49\textwidth]{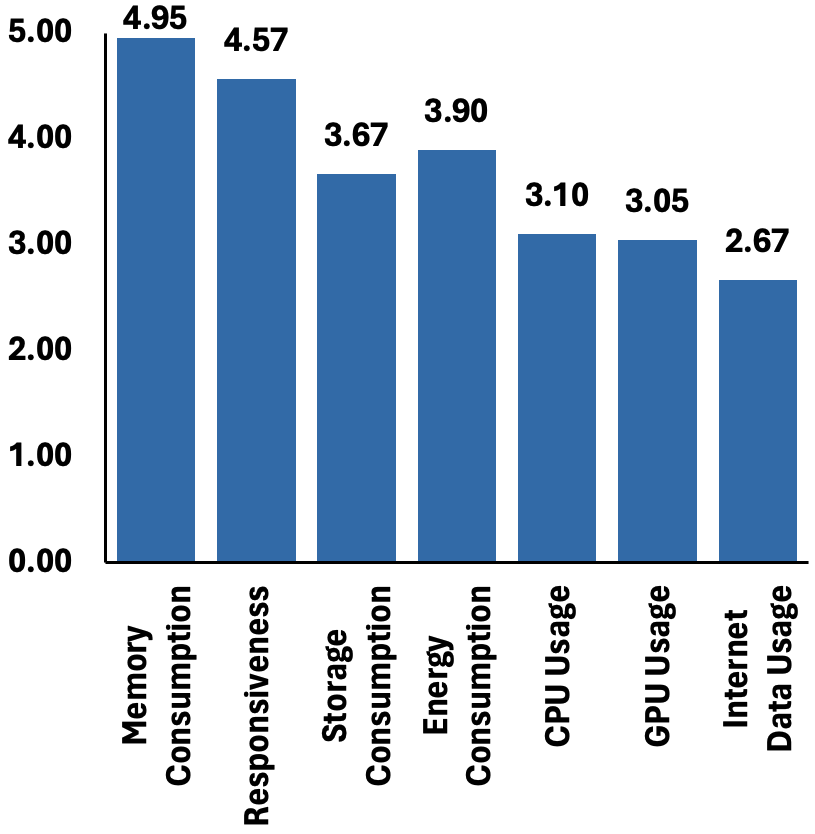}
    \caption{Average concern scores for different categories of Android performance issues.}
    \label{fig_average_concern_scores_for_PI}
\end{wrapfigure}

\subsubsection{Results}
We received 21 valid responses.
Participants’ ages fell into three ranges: 23 or below (4.8\%), 24–27 (76.2\%), and 28 or above (19.0\%), with the majority in the 24–27 range. 
Most participants (81\%) are casual developers with Android development experience, while the remaining 19\% are professional developers.
On average, participants reported 4.19 years of experience in Android development or related research.
For Q2, the median concern score across all seven performance issue categories was 4, indicating that participants widely recognize these issues as important in practice.
Fig.~\ref{fig_average_concern_scores_for_PI} shows the average concern level for each performance issue category.
The most concerning issue was \textit{Memory Consumption} (mean: 4.95), while the least concerning was \textit{Internet Data Usage}, which aligns with the patterns we observed in real-world discussions.
For Q3, participants selected an average of 54.19 out of 63 contributing factors, resulting in an overall alignment rate of 86.16\%.
Notably, 18 factors were selected by all participants, such as the \textit{UI Rendering} factor contributing to \textit{GPU Usage} issues and the \textit{Resource} factor contributing to \textit{Energy Consumption} problems.
This suggests strong agreement between our taxonomy and developers’ real-world experiences.
In contrast, some factors received fewer selections.
For example, the \textit{Third-party Library} factor contributing to \textit{CPU Usage} issues and the \textit{Singleton} factor contributing to \textit{Memory Consumption} issues were each selected by only 11 participants.
This may be because these factors occur less frequently or are less visible during development, leading to them being unintentionally overlooked by developers.

Overall, the survey results support the validity of our taxonomy and confirm that the performance issues and contributing factors identified in our study accurately reflect real-world development concerns, thereby reinforcing the findings presented in RQ1.

\section{Studies on Performance Issues}
\label{sec_STATUS_QUO_UNDERSTANDING}

In the above Section~\ref{sec_realworld_study}, we systematically examined real-world Android performance issues, their root causes, and common code patterns.
However, it remains unclear whether cutting-edge literature targets the same practical issues (RQ2.1), whether existing techniques can effectively resolve them (RQ2.2), and whether current datasets can comprehensively evaluate the techniques' effectiveness (RQ2.3).
To answer these questions, we conducted a systematic literature review (SLR) to assess the current state of Android performance analysis.
By comparing the performance issues, techniques, and datasets discussed in research papers with those observed in real-world discussions, we aim to uncover gaps between academic research and real-world challenges.

\subsection{Literature Review} \label{sec_literature_review}
\begin{figure*}
    \centering
    \setlength{\belowcaptionskip}{-10pt}
    \includegraphics[width=0.99\textwidth]{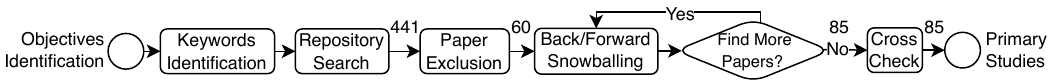}
    \caption{The working process of our systematic literature review.}
    \label{fig:literature review approach}

\end{figure*}

To conduct a systematic literature review, we followed the methodologies outlined by Brereton et al. ~\cite{brereton2007lessons} and Kitchenham et al. ~\cite{Kitchenham2004ProceduresFP}. 
Fig.~\ref{fig:literature review approach} illustrates the working process which comprises 5 steps.
Here, we clarify that this study focuses on performance issues resulting from inefficient coding practices within Android applications, rather than problems related to Android device hardware or the operating system.

\subsubsection{\textbf{Keywords Identification}}
To investigate the status quo of Android performance analysis, we identified a bunch of search keywords to locate relevant publications on this topic. 
Specifically, the keywords comprise two groups (i.e., G1 AND G2). 
Each group is made up of several keywords. 
G1 includes terms related to Android mobile phone-related keywords.
G2 includes terms related to performance issues identified in Section~\ref{sec_realworld_study}. 
Please note that to ensure comprehensive coverage and avoid missing relevant studies, we also included broader terms like ``performance'' and ``issue''.
The search string is then constructed by combining them, i.e., g1 AND g2, where g1 and g2 represent the disjunction of keywords from groups G1 and G2, respectively.
Please note that these keywords are based on our initial understanding of performance issues and are intended to help us identify ``seed'' papers.
As we review these papers, we could uncover new performance issues, prompting a snowballing process that could lead us to further relevant studies.
The * symbol represents any string of characters. 
For example, *phone* can refer to smartphone, phones, or phone.
Similarly, issue* can represent issues or issue.
Keywords are as follows:

G1: \emph{android, mobile, *phone*}

G2: \emph{resource, response, energy, memory, launch, freeze, lag, battery, responsiveness, cpu, gpu, storage, latency, performance, issue*}

\begin{wraptable}{R}{0.5\textwidth} 
\centering
\setlength{\belowcaptionskip}{0pt}
\setlength{\abovecaptionskip}{0pt}
\caption{CORE A/A* ranked software engineering and mobile venues.}
\label{tab:venues}
\resizebox{\linewidth}{!}{
\begin{tabular}{lllp{0.7\linewidth}}\hline
Type & Source & Research Codes & Venues  \\\hline

Journals & CORE2020 & 0803, 0805 & TSE, TOSEM, EMSE, JSS, IST, TMC  \\
Conferences & CORE2023 & 4606, 4612 & MOBICOM, Mobisys, EuroSys, MOBIHOC, MSWIM, ASE, ESEC/FSE, ICSE, EASE, ECSA, ESEM, FOSSACS, ICSA, ICSME, ICST, ISSRE, ISSTA, MSR, SANER, SEAMS  \\\hline

\end{tabular}}
\vspace{-5mm}
\end{wraptable}

\subsubsection{\textbf{Repository Search}} \label{sec_repo_search}
To ensure systematic literature collection, we conducted keyword-based searches across five libraries: 
\href{https://ieeexplore.ieee.org/search/advanced}{\emph{IEEE Library}}, \href{https://dl.acm.org/search/advanced}{\emph{ACM Library}}, \href{https://link.springer.com/}{\emph{Springer Library}}, \href{https://www.sciencedirect.com/search/entry}{\emph{Science Direct Library}}, and \href{https://onlinelibrary.wiley.com/search/advanced}{\emph{Wiley Library}}. 
From this initial search, we retrieved a total of 6,649 papers from IEEE, 1,507 from ACM, 537 from Springer, 1,969 from ScienceDirect, and 1,008 from Wiley.
Since many papers appeared across multiple libraries, we removed duplicates based on paper titles, resulting in 11,574 unique papers.
However, many of these studies were outside the scope of our focus. To narrow down to the most relevant papers, we concentrated on the fields of software engineering and mobile, selecting research codes 4606 and 4612 for conferences, and 0805 and 0803 for journals, specifically targeting publications containing the keywords ``software'' and ``mobile''.
In addition, we exclusively chose those published from conferences and journals rated CORE A*/A\footnote{CORE provides conference and journal ratings. For conference: \url{https://portal.core.edu.au/conf-ranks/}. For journal: \url{https://portal.core.edu.au/jnl-ranks/}} from 2012 to 2024. 
We then systematically reviewed each of these venues, applying the aforementioned keywords to search for the titles of relevant publications.
As a result, we successfully identified 441 publications distributed across 23 venues as listed in Table~\ref{tab:venues}. Notably, there is no relevant paper identified in ECSA, FOSSACS, and SEAMS.

\subsubsection{\textbf{Paper Exclusion}} \label{sec:paper exclusion}
To gather as many relevant papers as possible, we initially considered all returned results from our search.
However, not all papers aligned with our focus on performance issues in mobile apps caused by poor coding practices, as discussed in Section~\ref{sec:Intro}.
To refine our collection, we conducted a thorough review of the abstracts (and full content when necessary) of these papers, aiming to retain only those closely related to our research focus. 
Specifically, we applied the following exclusion criteria to streamline this process:
\begin{enumerate}[leftmargin=*]
    \item Papers targeting on non-Android devices (e.g., Windows Phone, IOS devices) were excluded. 
    \item Papers using Android performance issues as metrics/indicators for comparison were excluded. For instance, Mantis~\cite{Kwon2013MantisAP} predicts app execution time based on app implementations, and Eprof~\cite{Pathak2012WhereIT} is a profiler used to monitor energy consumption in apps.
    \item Papers targeting Android performance issues but not focused on the mobile app level were excluded. 
    For example, Hypnos~\cite{Jindal2013HypnosUA} is used for detecting sleep conflicts in smartphone device drivers, which is outside our scope and focused on mobile app issues stemming from poor coding practices by developers.
\end{enumerate}

After applying these exclusion criteria, we retained 60 papers that are closely related to the topic of performance issues analysis at the Android mobile level.

\subsubsection{\textbf{Back/Forward Snowballing}} \label{sec_snowball}
After initially identifying relevant papers (i.e., 60 papers), we employed a backward and forward snowballing method to ensure comprehensive coverage of closely related literature. 
The backward snowballing process involved scrutinizing the references of identified papers for additional relevant sources.
Additionally, we performed the forward snowballing method by examining papers citing the identified papers.
We strictly adhered to the exclusion criteria when conducting snowballing until no further papers could be identified. 
Eventually, we were able to locate a total of 85 papers.

\subsubsection{\textbf{Cross Check}}
Additionally, to avoid overlooking any primary studies, we conducted cross-checks involving at least two authors, ensuring the reliability of these papers and enhancing the trustworthiness of the primary studies.  
Ultimately, we collected 85 papers to assess the current state of research and identify gaps in addressing real-world performance issues.
As shown in Fig.~\ref{fig:publication_venue_type}, the majority of these studies were published in conference proceedings (62 out of 85, accounting for 72.94\%).
Moreover, we observe a consistent research interest in this topic, with papers published every year since 2012.
All papers involved in the SLR process are available in our replication package under the \textit{Literature\_Review/data} directory~\cite{our_repo}.

\subsection{Results}
In this section, we summarize the current state quo of Android performance analysis by thoroughly reviewing these selected papers. 
Our aim is to assess whether academic research objectives, tools, and datasets align with the real-world challenges faced by developers and users. Specifically, we evaluate:
1) Targeted performance issues and key factors that researchers focus on (Section~\ref{sec:problem}),
2) Automatic techniques designed to address Android performance issues (Section~\ref{sec_result_tech}), and
3) Datasets designed to evaluate the effectiveness of these automatic techniques (Section~\ref{sec_result_dataset}).

\subsubsection{\textbf{Result - Targeted Performance Issues}} \label{sec:problem}
In our literature review, we categorized the performance issues discussed in the collected papers and identified the contributing factors to each issue.
This approach allowed us to supplement the performance issues and factors not uncovered in our real-world exploration to obtain a holistic taxonomy.
Besides, by comparing these targeted issues and factors in academic research with those observed in real-world discussions, we evaluated how well researchers’ focus aligns with the practical challenges faced by developers and users.

\textbf{Holistic Taxonomy.}
Through reviewing 85 research papers, we identified five types of performance issues and 45 contributing factors, marking each in the \textit{Literature} column of Table~\ref{tab:performance__issues} with a checkmark (\ding{51}).
By incorporating insights from both real-world discussions and research papers, we developed a final taxonomy encompassing seven types of performance issues and 82 factors.
By examining these issues and their contributing factors from both real-world data and academic findings, we identified significant gaps between research focus and real-world challenges as follows.

\begin{figure}[htbp]
    \centering
    \begin{subfigure}[t]{0.48\textwidth}
        \centering
        \includegraphics[width=\textwidth]{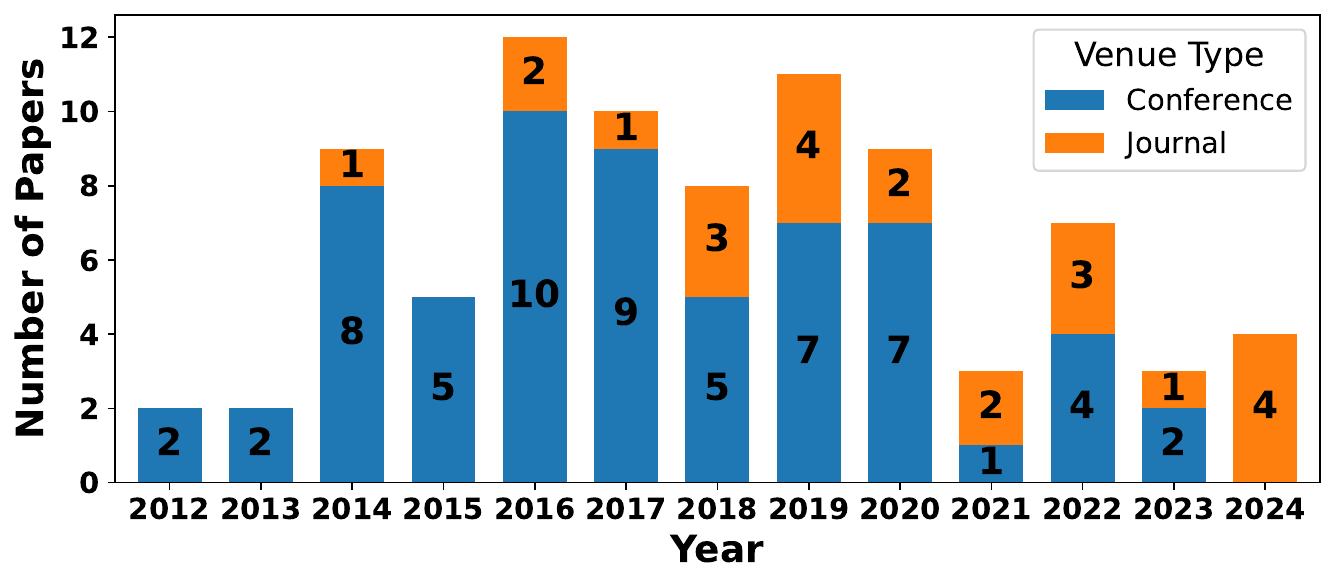}
        \caption{Number of publications per year by venue type.}
        \label{fig:publication_venue_type}
    \end{subfigure}
    \hfill
    \begin{subfigure}[t]{0.48\textwidth}
        \centering
        \includegraphics[width=\textwidth]{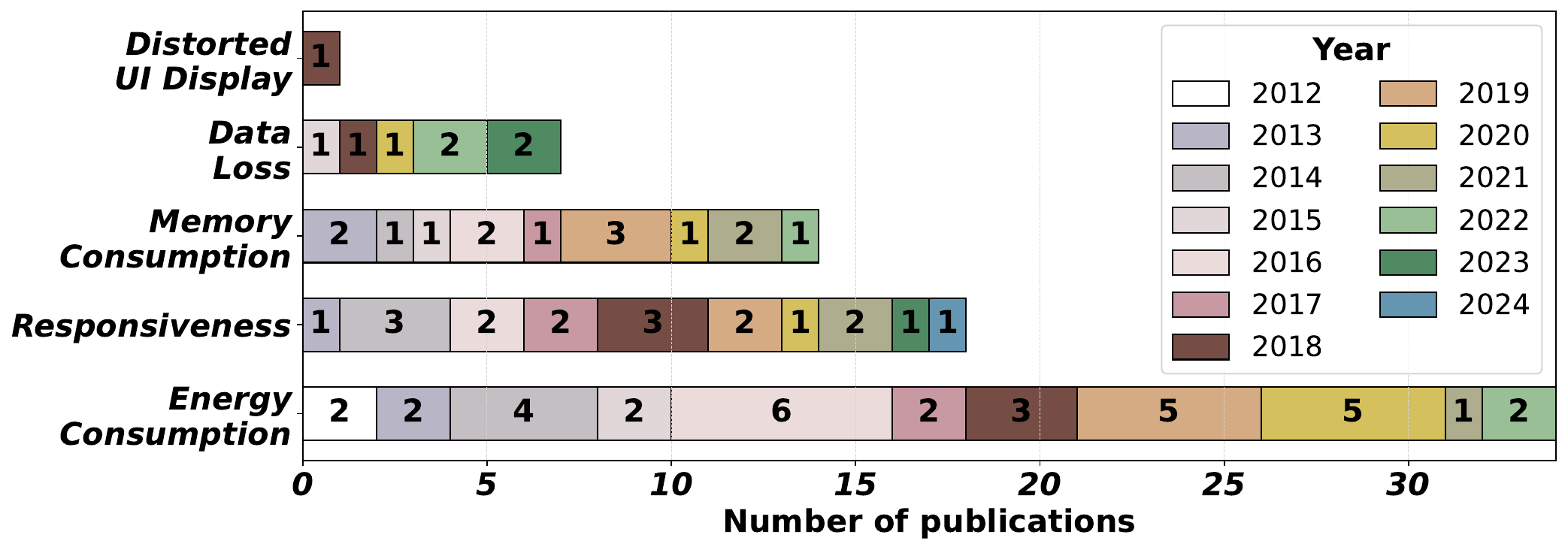}
        \caption{Annual publications on each performance issue.}
        \label{fig:publication_performance_year}
    \end{subfigure}
    \caption{Distribution of selected studies by venue type and performance issue.}
    \label{fig:combined_publication_stats}
\end{figure}

\textbf{Gaps between research focus and real-world performance issues.}
As shown in Fig.~\ref{fig:publication_performance_year}, we identified five key performance issues from literature: 
\textit{energy consumption}, \textit{responsiveness}, \textit{memory consumption}, \textit{GPU usage}, and \textit{CPU usage}.
These performance issues are also prevalent in real-world scenarios, indicating that researchers have made significant efforts to address concerns that are relevant to both developers and users.
Despite these efforts, there remains a noticeable imbalance in focus.
Among the five issues, energy consumption received the most attention, with 69 of the examined papers (81.18\%) focusing on this issue, and this trend is likely to continue.
However, despite this focus on energy consumption, Section~\ref{sec:realworld data analysis} revealed that users are more concerned with responsiveness, while developers prioritize memory consumption.
This highlights a discrepancy between the performance issues prioritized by researchers, developers, and users.
Moreover, two performance issues that are prevalent in real-world applications, \textit{storage consumption} and \textit{internet data usage}, have been relatively underexplored in the academic literature~\cite{Bijlani2021WhereDM,Zhang2012HowEA,Liu2016WhoMM}, highlighting the need for more focused attention in future research.
These gaps suggest that researchers should place greater emphasis on the performance issues that are of primary concern to both users and developers in real-world apps.

\textbf{Gaps in academic research on real-world performance issue factors.}
By analyzing the cases where both the \textit{Real-World} and \textit{Literature} columns are marked with a checkmark (\ding{51}), we found that only 27 out of 63 factors discussed in real-world settings (42.86\%) have been studied in literature.
It indicates that the majority of performance issues encountered in real-world scenarios remain underexplored in academic research.
For instance, \textit{memory consumption}, which is the top concern for developers, real-world data shows 21 contributing factors, but only 8 (38.10\%) have been explored in literature. 
Similarly, for \textit{responsiveness}, the top concern for users, 16 real-world factors were identified, but only 10 (62.50\%) have been studied in literature.
Less common issues like \textit{CPU usage} and \textit{GPU usage} have seen even less attention, with only 14.29\% and 50.00\% of their respective real-world root causes being studied in literature.
These gaps suggest that many root causes of real-world performance issues still remain unexplored by researchers. 
Notably, all identified root causes of \textit{energy consumption} have been addressed in literature, indicating that this area has been extensively studied.
This observation aligns with our finding that the majority of research efforts are concentrated on energy consumption.

\textbf{The presence of research-only factors not seen in real-world discussions.}
By examining rows where the \textit{Real-World} column is unmarked (\ding{55}) but the \textit{Literature} column is marked (\ding{51}), we identified 19 factors that have been explored in literature but not observed in the real world. 
Specifically, we found one unseen factor in \textit{memory consumption}, four in \textit{responsiveness}, and 14 in \textit{energy consumption}.
This suggests that literature may uncover factors that did not appear in our real-world exploration, thereby further enriching the taxonomy.
Notably, most of these unseen factors pertain to \textit{energy consumption}, accounting for 73.68\% of the factors explored in literature but not observed in the real-world.
For example, Li et al.~\cite{Li2014MakingWA} found that OLED displays consume more energy when displaying light colors than dark colors. 
Similarly, Cito et al.~\cite{Cito2016BatteryawareTI} observed that frequent advertisement and analytics (A\&A) service requests lead to significant energy waste.
On the other hand, there is a substantial number of energy consumption factors that appear only in the literature, yet rarely manifest in real-world applications. This indicates a pronounced disconnection between academic research and the practical challenges faced by practitioners.

\find{\textbf{Answers to RQ2.1:}
\begin{itemize} [leftmargin=*]
    \item 
    Through an extensive integration of real-world explorations and thorough literature review, we present a comprehensive taxonomy encapsulating seven primary types of performance issues and 82 contributing factors.
    \item A noteworthy finding shows that only 42.86\% of factors contributing to performance issues have been examined in academic research.
    \item A significant misalignment exists in the primary concerns of researchers, developers, and users: researchers concentrate on energy consumption, developers prioritize memory usage, and users are chiefly concerned with responsiveness.
\end{itemize}
}

\subsubsection{\textbf{Result - Technique}} \label{sec_result_tech}
To address performance issues, researchers have proposed automatic tools using diverse techniques, categorized into static approaches, dynamic approaches, and deep learning approaches.
Table~\ref{tab:techniques} presents the list of available automatic approaches collected from our examined papers.
Among the 85 papers examined, 14 (16.47\%) provided publicly accessible tools designed to detect specific factors contributing to particular performance issues.
This small fraction can be attributed to several reasons:
some papers lack accessible artifacts, others provide only datasets without the corresponding tool code, and in some cases, the links to the artifacts are invalid.
This indicates a significant gap between automatic tools and their practical usability within the community.

In addition, we found that existing tools target only 30 out of the 82 factors in our taxonomy (Table~\ref{tab:performance__issues}), covering just 36.59\%.
This indicates that the majority of factors contributing to the seven primary types of performance issues are overlooked by SOTA tools, highlighting a significant gap in addressing these issues comprehensively.
Moreover, most automated tools are primarily designed to detect performance issues rather than actively resolve them. Given that these detection tools rely on specific code patterns, we can anticipate that the identified patterns can also be applied to remediate performance issues.
We then summarize the approaches as follows:

\begin{table}[!htbp]
\centering
\caption{A summary of automatic approaches used for Android mobile performance analysis. \textit{Static} denotes Static Analysis, \textit{Dynamic} denotes Dynamic Analysis, and \textit{DL} denotes Deep Learning methods.}
\label{tab:techniques}
\resizebox{0.85 \textwidth}{!}{
\begin{tabular}{l|c|c|c|c|c|c} 
\hline
Tool & Approach & Technique & Stage & \makecell[c]{Performance Issue} & \makecell[c]{Target Factor} & Link \\ 
\hline

VALA~\cite{Lu2022DetectingRU} & Static & \makecell[c]{Dataflow Analysis\\Symbolic Execution} & Detect  & \makecell[c]{Memory Consumption\\Energy Consumption} & \makecell[c]{Activity\\Resource} & ~\cite{Lu2022DetectingRUTool} \\
\hline

\small AsyncChecker\cite{Pan2020StaticAC}   & Static &  \makecell[c]{Code Pattern Analysis\\Interprocedural- \\ Code Analysis} & Detect  & \makecell[c]{Memory Consumption\\Energy Consumption\\Responsiveness} & AsyncTask & ~\cite{Pan2020StaticACTool} \\
\hline

TAPIR~\cite{Li2019CharacterizingAD}  & Static & \makecell[c]{Code Pattern Analysis}  & Detect  & \makecell[c]{Memory Consumption\\Energy Consumption\\Responsiveness} & Image & ~\cite{Li2019CharacterizingADTool} \\
\hline

ServDroid~\cite{Song2019ServDroidDS}   & Static & \makecell[c]{Code Pattern Analysis\\Interprocedural Code Analysis}  & Detect   & \makecell[c]{Memory Consumption\\Energy Consumption} & Service & ~\cite{Song2019ServDroidDSTool} \\
\hline

Lyu et al.~\cite{Lyu2018RemoveRF}  & Static &  \makecell[c]{Dataflow Analysis\\Program Transformation}   & \makecell[c]{Detect\\Fix}  & \makecell[c]{Energy Consumption\\Responsiveness} & \makecell[c]{Database Operation\\Loop}  & ~\cite{Lyu2018RemoveRFTool} \\
\hline

PlumbDroid~\cite{Bhatt2020AutomatedRO} & Static & \makecell[c]{Control-Flow Analysis\\Patch Generation} & \makecell[c]{Detect\\Fix} &  Responsiveness &
Resource & ~\cite{Bhatt2020AutomatedROTool} \\ \hline

\textmu Droid~\cite{Jabbarvand2017DroidAE} & Dynamic & \makecell[c]{Mutation Testing\\Power Trace Analysis} &  Detect &  Energy Consumption & \makecell[c]{Resource\\Wakelock\\Loop} & ~\cite{Jabbarvand2017DroidAETool} \\ \hline

ACETON~\cite{Jabbarvand2020AutomatedCO}  & \makecell[c]{DL \\Dynamic} & \makecell[c]{Deep Learning\\Feature Engineering\\Runtime Monitoring}  & Detect  & Energy Consumption & Resource & ~\cite{Jabbarvand2020AutomatedCOTool} \\
\hline

Leafactor~\cite{Cruz2017LeafactorIE} & Static & \makecell[c]{Code Pattern Analysis\\Program Transformation} & \makecell[c]{Detect\\Fix} & Energy Consumption & \makecell[c]{Wakelock\\View\\UI Rendering} & ~\cite{LeafactorTool} \\ \hline

PAPRIKA~\cite{Hecht2015DetectingAI} & Static & \makecell[c]{Code Pattern Analysis} &  Detect & \makecell[c]{Memory Consumption\\Responsiveness\\Energy Consumption}  &  \makecell[c]{Inefficient Method\\Data Type\\Activity} & ~\cite{PAPRIKATool} \\ \hline

ecoCode~\cite{LeGoaer2022ecoCodeAS} & Static & \makecell[c]{Rule-Based Detection\\XML/Gradle Inspection} & Detect & Energy Consumption & \makecell[c]{Resource\\Wakelock\\Redundant Frames\\Service\\Loop\\HTTP Request} & ~\cite{ecoCodeTool} \\ \hline

aDoctor~\cite{Iannone2020RefactoringAE} & \makecell[c]{Static} & \makecell[c]{Code Pattern Analysis\\Program Transformation} & \makecell[c]{Detect\\Fix} & Energy Consumption & \makecell[c]{Wakelock\\Inefficient Method\\Data Type} & ~\cite{adoctorTool} \\ \hline

Das et al.~\cite{Das2020CharacterizingTE} & \makecell[c]{Static} & \makecell[c]{Rule-Based Detection\\AST Analysis} & Detect & \makecell[c]{Energy Consumption\\Memory Consumption\\Responsiveness} & \makecell[c]{UI Rendering\\Data Type\\Main Thread\\Inefficient Method\\View\\Wakelock\\Activity} & ~\cite{AndroidLintTool} \\ \hline

Vásquez et al.~\cite{Vsquez2015HowDD} & \makecell[c]{Dynamic} & \makecell[c]{Runtime Monitoring} & Detect & \makecell[c]{Responsiveness} & \makecell[c]{Main Thread} & ~\cite{StrictModeTool} \\ \hline

\end{tabular}
}
\end{table}

\textbf{Static Analysis.}
Most existing approaches are based on static analysis to detect or fix performance issues, as shown in Table~\ref{tab:techniques}.
This is intuitive since static analysis involves reasoning about program behavior without execution, making it lightweight to deploy.
Specifically, taint analysis, symbolic execution~\cite{Lu2022DetectingRU}, code pattern analysis~\cite{Pan2020StaticAC, Li2019CharacterizingAD, Song2019ServDroidDS}, and data flow analysis~\cite{Lyu2018RemoveRF} are commonly employed in rule-based static analysis as many performance issues arise from common code patterns. 
Additionally, static approaches, such as feature engineering, are used to assist deep learning approaches by extracting semantic code features~\cite{Kim2019AMD,Su2016ADL}. 
Moreover, they can also provide fundamental information for dynamic fuzzing tools to trigger performance bugs on the fly~\cite{Guo2022DetectingAF}. 
Furthermore, static approaches play an important role in fixing performance issues, particularly in generating patches for automatic program repair~\cite{Lyu2018RemoveRF, Bhatt2020AutomatedRO}.

\textbf{Dynamic Analysis.}
There are only three studies that primarily employ dynamic analysis approaches to detect performance bugs. This result aligns with expectations because dynamic analysis involves executing apps in a controlled environment and is unable to cover all possible paths~\cite{dong2020timemachine}. However, dynamic analysis is more resilient to evasion techniques (e.g., reflection~\cite{sun2021taming}, obfuscation~\cite{sun2023demystifying}) and can supplement static approaches to eliminate false positives~\cite{sun2022mining}. 
As shown in Table~\ref{tab:techniques}, two dynamic analysis tools target energy consumption issues, particularly unnecessary energy waste such as resource under-utilization, wakelock misuse, and inefficient loops, which are detectable through abnormal energy consumption monitoring.
One dynamic analysis tool, StrictMode~\cite{StrictModeTool}, targets responsiveness issues, particularly those caused by executing heavy operations on the main thread, which can lead to UI freezes and slow interactions.
We also discovered one study~\cite{Jabbarvand2020AutomatedCO} that employed dynamic monitoring to collect features for deep learning methodologies aimed at detecting energy issues caused by resource inefficiencies.

\textbf{Deep Learning.}
Despite the growing popularity of deep learning (DL) in software engineering tasks like code analysis~\cite{Allamanis2017LearningTR}, bug detection~\cite{Pradel2018DeepBugsAL}, and program repair~\cite{zhao4254659autopatch}, surprisingly, only one study applied DL to Android performance analysis. 
Specifically, Jabbarvand et al.~\cite{Jabbarvand2020AutomatedCO} observed runtime features dynamically and leveraged DL models to locate energy anomalies.
The limited adoption of applying DL to Android performance analysis may due to challenges in collecting large datasets that are sufficient to train DL models on performance issues and their corresponding fixes.

\find{\textbf{Answers to RQ2.2:} 
\begin{itemize} [leftmargin=*]
    \item We identified 14 publicly available tools, primarily using static analysis, dynamic analysis, and deep learning approaches.
    \item Existing tools target only 30 of the 82 identified factors to performance issues (36.59\%), leaving a substantial number of factors unexamined.
    \item Most tools focus on detecting performance issues while neglecting their resolution.

\end{itemize}
}

\begin{table}[!htbp]
\centering

\caption{A list of datasets used for Android mobile performance analysis, based on our literature review.}
\resizebox{0.85 \textwidth}{!}{
\begin{tabular}{l|c|c|c|c|c|c} 
\hline
Dataset & Year &\# APP & Type & Performance Issue & Target Factor & Link \\ 
\hline

Lyu et al.~\cite{Lyu2018RemoveRF} & 2018  & 206 &  Real-World & \makecell[c]{Energy Consumption\\Responsiveness} & \makecell[c]{Database Operations\\Loop} & ~\cite{Lyu2018RemoveRFTool} \\ \hline 

V{\'a}squez et al.~\cite{LinaresVsquez2018MultiObjectiveOO} & 2018 & 5 &  Real-World & \makecell[c]{Energy Consumption} & \makecell[c]{Display Screen} & ~\cite{LinaresVsquez2018MultiObjectiveOOTool} \\ \hline 

Habchi et al.~\cite{Habchi2019TheRO} & 2019 & 324 &  Real-World & \makecell[c]{Memory Consumption\\Energy Consumption\\Responsiveness} & \makecell[c]{UI Rendering\\View\\Inefficient Method} & ~\cite{Habchi2019TheROData} \\ \hline 

Li et al.~\cite{Li2019CharacterizingAD} & 2019 & 19 &  Real-World & \makecell[c]{Memory Consumption\\Energy Consumption\\Responsiveness} & Image & ~\cite{Li2019CharacterizingADTool} \\ \hline

AsyncBench~\cite{Pan2020StaticAC} & 2020 & 69 &  Micro & \makecell[c]{Memory Consumption\\Energy Consumption\\Responsiveness} & AsyncTask & ~\cite{Pan2020StaticACTool} \\ \hline 

Liu et al.~\cite{Liu2019DroidLeaksAC} & 2020 & 34 &  Real-World & \makecell[c]{Memory Consumption\\Energy Consumption} & \makecell[c]{Resource} & ~\cite{Liu2019DroidLeaksACTool} \\ \hline 

Das et al.~\cite{Das2020CharacterizingTE} & 2020 &  316 &  Real-World & \makecell[c]{Memory Consumption\\Energy Consumption} & \makecell[c]{UI Rendering\\Data Type\\Inefficient Method\\View\\Wakelock} & ~\cite{Das2020CharacterizingTETool} \\ \hline

Lu et al.~\cite{Lu2022DetectingRU} & 2022 & 35 &  Real-World & \makecell[c]{Memory Consumption\\Energy Consumption} & \makecell[c]{Resource} & ~\cite{Lu2022DetectingRUTool} \\ \hline 

Janssen et al.~\cite{Janssen2022OnTI} & 2022 & 40 &  Real-World & \makecell[c]{Energy Consumption\\Responsiveness} & \makecell[c]{UI Rendering} & ~\cite{Janssen2022OnTITool} \\ \hline 

Petalotis et al.~\cite{Petalotis2024AnES} & 2024 & 9 &  Real-World & \makecell[c]{Energy Consumption\\Responsiveness} & \makecell[c]{Advertisements} & ~\cite{Petalotis2024AnESTool} \\ \hline

Cruz et al.~\cite{Cruz2019CatalogOE} & 2019 & 1,027 &  Real-World & Energy Consumption & \makecell[c]{Display Screen\\UI Rendering\\Wakelock\\Resource\\Obsolete Task} & ~\cite{Cruz2019CatalogOEDataset} \\ \hline

Prestat et al.~\cite{Prestat2024DynAMICSAT} & 2024 & 5 & Real-World  & \makecell[c]{Energy Consumption\\Responsiveness\\Memory Consumption} & \makecell[c]{Wakelock\\Inefficient Method\\Resource\\AsyncTask\\View} & ~\cite{Prestat2024DynAMICSATDataset} \\ \hline

\end{tabular}
}
\label{tab:datasets}

\end{table}

\subsubsection{\textbf{Result - Dataset}} \label{sec_result_dataset}

To collect datasets, several types of papers need to be included:
1) Tool papers that construct datasets to evaluate their designed tools;
2) Empirical studies that collect specific datasets to analyze performance issues, where labeled data can be used to support evaluation.
After examining all the public links or websites within the 85 identified studies, we identified 12 (14.11\%) unique, available, and open-sourced datasets, as shown in Table~\ref{tab:datasets}. Notably, we conducted the examination in Jun 2025.
Since most of these datasets are designed for diverse performance factors, it becomes infeasible to ensure a fair comparison.
In addition, the number of factors covered by these datasets is quite limited. 
Out of the 82 factors identified in our taxonomy, only 24 (29.27\%) are addressed by existing datasets.
This indicates a significant need for constructing more comprehensive datasets.
Moreover, there is a considerable imbalance in dataset size.
The smallest dataset contains only five apps, while the largest has 1,027, with an average of 174.1 apps per dataset.
This imbalance makes it unrealistic for any single dataset to cover all potential issues comprehensively.
Even if some tools are effective at detecting performance issues, they may still face soundness challenges due to false negatives. 
Therefore, more effort is needed to create datasets that are not only comprehensive but also more representative of the diverse range of performance issues.

\find{\textbf{Answers to RQ2.3:}
 
\begin{itemize} [leftmargin=*]
    \item We identified 12 open-source datasets for evaluating performance issues; however, they encompass only 24 out of 82 factors (29.27\%), highlighting a significant gap in tool evaluation.
    \item  These datasets suffer from inadequate sample sizes, with the smallest containing only five apps and the largest consisting of 1,027.
\end{itemize}
}

\section{Future Trends} \label{sec_future_trends}

Through our real-world exploration (Section~\ref{sec_realworld_study}) and literature review (Section~\ref{sec_STATUS_QUO_UNDERSTANDING}), we have revealed significant gaps in current academic research:
it addresses only a fraction of the performance issues that developers and users prioritize, existing tools lack the capacity to comprehensively tackle these issues, and current datasets are insufficient for effectively evaluating tool performance.
Drawing upon these findings, we present discussion points regarding the research and practice of Android performance analysis, aiming to guide future researchers in this field.

\subsection{Targeted Objective Imbalance}
In Section~\ref{sec:problem}, we found that there is a misalignment in the primary concerns of researchers, developers, and users: researchers primarily focus on energy consumption, developers prioritize memory consumption, while users are most concerned with responsiveness.
This suggests that future research should increase focus on responsiveness and memory issues, as these have a more immediate impact on both end-users and developers.
We also found two categories of performance issues, \textit{storage consumption} and \textit{internet data usage}, that remain unexplored in academic research, indicating potential areas for further investigation.
Additionally, only 42.86\% of the factors contributing to performance issues observed in real-world settings have been explored in academic research, remaining nearly 60\% still awaiting for researchers to investigate.
Interestingly, there are 19 factors among the 82 factors that lead to performance issues that have only been explored in literature but not observed in the real-world scenarios.
This suggests that the issues currently discussed in academic research may not be of real concern to users or developers in real-world settings.
Overall, this imbalance between academic research objectives and real-world challenges underscores the need for greater focus on the real-world performance issues and contributing factors that are most relevant to end-users and developers.

\subsection{Automatic Performance Analysis Tools are Still in an Early Stage} \label{sec_future_tech}

Our study revealed that only 30 out of 82 factors (36.59\%) are addressed by state-of-the-art (SOTA) tools, with most tools focusing on detection rather than resolution.
In fact, only four out of the existing tools are dedicated to fixing performance issues. Although static, dynamic, and deep learning (DL) techniques have shown effectiveness in research experiments, applying these methods to holistically detect and fix performance issues in practice remains an unsolved challenge.
Future research should focus not only on building comprehensive frameworks to detect real-world performance issues, but also on providing solutions for fixing these issues.

The lack of variety in technique selection hinders the soundness, as static approaches are not proficient in handling complicated code features (e.g., reflection~\cite{sun2021taming}, obfuscation~\cite{samhi2022jucify}, and hardening~\cite{Schloegel2022LokiHC}), and it's non-trivial for dynamic tools to achieve satisfactory results to trigger sufficient paths in practice. Thus, developing effective tools against performance issues will be a noteworthy research topic in the future. 
In fact, DL has made considerable achievements in software engineering~\cite{Wang2023MachineDeepLF,Feng2020CodeBERTAP,Gu2016DeepAL,Hu2018DeepCC,pan2024don}, and these advanced techniques have demonstrated powerful capabilities in solving many complex tasks (i.e., code analysis~\cite{White2016DeepLC,Jebnoun2021ClonesID,Zhao2018DeepSimDL} and code generation~\cite{Wang2022TestDrivenML, zhang2025deployability}). 
Applying advanced DL techniques to detect and fix Android performance issues is promising, although significant challenges remain in collecting a sufficient performance issue dataset for model training.
The emergence of large language models (LLMs) also presents new possibilities.
LLMs, with their in-context learning capabilities, could address the limitations of traditional DL approaches by enabling models to detect and resolve performance issues with fewer examples and greater adaptability~\cite{Hou2023LargeLM, Liao2023A3CodGenAR, Si2024AST, Fan2023LargeLM, pan2024large, han2024chase, Huang2022SEFK, Su2024EnhancingET}.
This opens an exciting direction for future research in Android performance analysis, promising to bridge the current gaps in existing techniques.

\subsection{Constructing Benchmark Dataset is an Urgent Need} \label{sec_future_dataset}
Similar to the shortage of tools, current datasets are too sporadic to reflect real performance issues, evidenced by the fact that only 24 (29.27\%) of the 82 factors are covered by existing datasets. 
Even among the 24 factors, the data distribution is highly imbalanced, as the smallest dataset contains only five apps, while the largest has 1,027. 
In addition, the amount of available data in each dataset is very small (usually only a few dozen examples), which is insufficient to support a fair evaluation of automated tools. 
Another critical issue is reproducibility.
Our study indicates that only a small number of primary studies share valid datasets, making it difficult for future researchers to validate and evaluate tools. Furthermore, these datasets lack real-time updates for the latest app, thus containing outdated samples.
Thus, it is urgent to construct up-to-date, comprehensive benchmark datasets to support fair comparisons, covering as many real-world cases as possible.

\section{Threats to Validity} \label{sec_threats_to_validity}
Our study is subject to several potential threats to validity, which we categorize into internal and external aspects.

\subsection{Internal Validity}
This concerns the soundness of our data collection and literature review processes.
For \textbf{real-world data filtering}, the process may miss some relevant information. 
This is because the sentiment classification model~\cite{hartmann2023} used to filter user reviews may not guarantee 100\% accuracy in identifying negative reviews. 
Additionally, the Google Play scraper might fail to capture all user reviews. 
Lastly, there could be biases in our manual checking process.
To minimize subjective differences, two authors independently checked the filtered data. 
In cases of disagreement, they discussed and resolved whether to include the data.
For \textbf{literature review}, despite using a well-established methodology~\cite{Kitchenham2004ProceduresFP}, we cannot guarantee that our empirical results cover all relevant works. 
To locate relevant studies, we formulated search strings from both Android and performance issues, as described in Section~\ref{sec_repo_search}, and used five well-known electronic databases. 
To minimize the risk of missing relevant work, we conducted backward and forward snowballing and cross-checked our results.

\subsection{External Validity}
This concerns the generalizability of our findings beyond the studied datasets and contexts.
Regarding the \textbf{representativeness of data sources}, our real-world data was derived from open-source and community-driven platforms such as GitHub, Stack Overflow, and Google Play. While these platforms reflect commonly reported issues from general developers and users, they may not capture the performance concerns of proprietary or enterprise-level applications developed by organizations such as Google or Meta. Furthermore, Stack Overflow discussions may be biased toward novice or intermediate developers, as experienced professionals may participate less frequently.
As for the \textbf{research scope}, this study focuses exclusively on performance issues in Android \textit{mobile} applications. Other Android-based platforms, such as Android TV, Android Auto, or IoT systems, may exhibit different performance behaviors due to variations in hardware, user interactions, and platform-specific constraints. These platforms lie beyond the scope of this work and represent promising avenues for future research.
Another external threat lies in the lack of \textbf{user-centric validation}. 
While our developer survey provides evidence that the taxonomy reflects practical development concerns, we did not conduct a complementary study with end users. 
Consequently, our taxonomy may not fully capture performance issues that users subjectively perceive as most critical. 
Future work could incorporate user-centric validation, for example through controlled user studies or large-scale questionnaires, to further strengthen the external validity of our findings.
\section{Related Work}

To investigate Android performance issues, numerous studies have been conducted, which can be broadly categorized into two directions:  
(1) empirical studies based on real-world data sources, and  
(2) literature reviews that summarize the current state of performance-related research.

\subsection{Empirical Studies on Android Performance Issues}

Numerous studies have explored Android performance issues using real-world sources~\cite{Kumari2024AnES, Noor2022EndUP, Das2016AQA, Vsquez2015HowDD, Liu2014CharacterizingAD, Rua2023ALE, Das2020CharacterizingTE, Habchi2019TheRO, Habchi2020AndroidCS, MazueraRozo2020InvestigatingTA, Prestat2022AnES}.  
These studies typically focus on a single type of data source and adopt either the user or developer perspective.
For example, Noor et al.~\cite{Noor2022EndUP} focused on the user perspective by examining 368,704 low-rated Google Play reviews from 55 apps. They identified eight categories of performance issues reported by users, including responsiveness delays, graphical glitches, and image rendering problems.
Kumari et al.~\cite{Kumari2024AnES} analyzed 385 Stack Overflow posts to understand performance issues discussed by developers.
They proposed two taxonomies: one categorizing eight performance issue types (e.g., UI/UX, graphics, and development practices), and another describing their associated causes.  
Similarly, Das et al.~\cite{Das2016AQA}, Vásquez et al.~\cite{Vsquez2015HowDD} and Liu et al.~\cite{Liu2014CharacterizingAD} analyzed GitHub issue reports and commit histories to categorize performance problems that developers frequently encounter.

While these studies effectively identify performance issues and root causes from user and developer perspectives, they do not assess whether these real-world concerns are adequately addressed by academic research.  
Moreover, although some works propose taxonomies, the classification dimensions and terminology vary significantly, hindering comparison and integration across studies.
In contrast, our study integrates data from multiple sources, including Google Play (user perspective), Stack Overflow and GitHub (developer perspective), and academic literature (researcher perspective).  
Building on this integration, we adopt a hierarchical consequence–cause structure that explicitly links observable performance issues to their underlying root causes, enabling direct mapping to existing tools and datasets for comprehensive gap analysis.

\subsection{Literature Reviews on Android Performance Issues}

To assess the current state of Android performance analysis, several systematic literature reviews have summarized the performance-related issues studied in academic research~\cite{Wu2023ASL, Fawad2024AndroidSC, Hort2021ASO}.  
For example, Wu et al.~\cite{Wu2023ASL} reviewed 35 studies on Android code smells and identified five major categories, including performance-related smells such as energy inefficiency and memory usage.  
Similarly, Fawad et al.~\cite{Fawad2024AndroidSC} analyzed 79 papers and extracted 237 types of Android code smells, covering concerns such as durable wakelocks, member-ignoring methods, and inefficient database access, many of which are related to performance.  
In addition, Hort et al.~\cite{Hort2021ASO} reviewed 156 papers on non-functional performance optimization in Android apps. These studies focused on four key performance characteristics (e.g., responsiveness, launch time, memory consumption, and energy usage), and summarized common optimization strategies such as anti-pattern removal, refactoring, and preloading.

While these reviews offer valuable insights into performance issues and proposed solutions, they do not evaluate whether academic efforts effectively address the most critical challenges in real-world development.  
They also overlook the extent to which existing tools and datasets support the detection and evaluation of such issues.  
Consequently, the actual alignment between academic research and practical concerns remains unclear.
In contrast, our study first identifies the full spectrum of real-world performance issues.  
We then analyze how many of these issues are examined in the literature, which tools are available to address them, and whether corresponding datasets exist to support empirical evaluation.  
This enables a comprehensive understanding of research coverage and highlights gaps to guide future work.

\section{Conclusion} \label{sec_conclusion}
In this paper, we conducted a comparative study to examine real-world Android performance issues encountered by users and developers, and to evaluate the extent to which current academic research addresses these practical challenges.
By analyzing large-scale discussions from three popular crowd-sourced platforms on Android performance issues, alongside a review of 85 high-quality research papers, we identified several gaps between real-world challenges and the focus of academic research.
Our main findings are as follows:
(1) A taxonomy categorizes seven types of performance issues and 82 contributing factors.
(2) Six common code patterns of these performance issues.
(3) There is a misalignment in the primary concerns of researchers, developers, and users.
(4) Only 42.86\% of the real-world factors have been explored in academic research.
(5) The majority of existing tools primarily utilize static analysis, dynamic analysis, and deep learning, with most tools detecting issues rather than fixing them.
(6) Existing available tools target only 36.59\% of identified factors.
(7) Existing available datasets cover only 29.27\% of factors and lack sufficient data to support effective tool evaluation.
Finally, we proposed future research directions to bridge the identified gaps.



\bibliographystyle{unsrt}
\bibliography{sample-base}

\end{document}